\documentclass[fleqn,usenatbib]{mnras}

\usepackage{newtxtext,newtxmath}
\usepackage[normalem]{ulem}
\usepackage{listings}
\usepackage{xspace} 
\usepackage[dvipsnames]{xcolor}

\usepackage[T1]{fontenc}

\DeclareRobustCommand{\VAN}[3]{#2}
\let\VANthebibliography\thebibliography
\def\thebibliography{\DeclareRobustCommand{\VAN}[3]{##3}\VANthebibliography}

\usepackage{graphicx}	
\usepackage{amsmath}	
\usepackage{comment}
\usepackage{color}

\newcommand{\St}{\mathrm{St}}

\newcommand{\dustpy}{\texttt{DustPy}\xspace}
\newcommand{\twopop}{\texttt{two-pop-py}\xspace}
\newcommand{\tripod}{\texttt{TriPoD}\xspace}
\newcommand{\cudisc}{\texttt{cuDisc}\xspace}
\newcommand{\mcdust}{\texttt{mcdust}\xspace}

\title[Turning the knobs on dust evolution]{Turning the knobs on dust evolution: Comparing codes, parameters and their effects on planet formation and disc observables} 

\author[Eriksson et al.]{
Linn E. J.~Eriksson$^{1}$\thanks{E-mail: leriksson@amnh.org},
Thomas Pfeil$^{2}$, Nicolas Kaufmann$^{3}$, and Vignesh Vaikundaraman$^{4}$
\\
$^{1}$ Department of Astrophysics, American Museum of Natural History, 200 Central Park West, New York, NY 10024, USA \\
$^{2}$ Center for Computational Astrophysics, Flatiron Institute, 162 5th Avenue, New York, NY 10010, USA\thanks{The Flatiron Institute is a division of the Simons Foundation} \\ 
$^{3}$ University Observatory, Faculty of Physics, Ludwig-Maximilians-Universität München, Scheinerstr. 1, 81679 Munich, Germany\\
$^{4}$ Max Planck Institute for Solar System Research, Justus-von-Liebig-Weg 3 37077 Göttingen
}

\date{Accepted XXX. Received YYY; in original form ZZZ}

\pubyear{2026}

\begin{document}
\label{firstpage}
\pagerange{\pageref{firstpage}--\pageref{lastpage}}
\maketitle


\begin{abstract}
Protoplanetary discs contain a wide range of dust sizes that influence their thermal structure and planet formation processes such as planetesimal formation and pebble accretion. Dust evolution models are therefore essential for both planet formation simulations and the interpretation of disc observations. Several open-source dust evolution codes are available, each adopting different model assumptions. We present a comparison of 1D radial simulations using (in order of complexity) \twopop{}, \tripod{}, and \dustpy{}, and 2D radial–vertical simulations with \tripod{}, \cudisc{}, and \mcdust{}. The comparison includes dust size distributions, dust disc masses, planetary gap structures, millimetre fluxes and disc sizes from synthetic observations, planetesimal formation regions, and planetary growth via pebble accretion. We also perform a parameter study to assess how key dust-evolution parameters influence disc evolution, planet formation, and code agreement. In 1D, \twopop{} depletes dust masses faster and produces higher dust concentrations outside planetary gaps than \dustpy{} or \tripod{}. The latter two generally agree well, except when size distributions deviate strongly from a power law. While the calculated millimetre fluxes and disc radii agree well, planetesimal formation locations and pebble accretion rates vary significantly between codes. 
In 2D, we compare \cudisc{}, \mcdust{}, and \tripod{} in simulations of turbulence- and sedimentation-driven coagulation. The dust size distributions agree well, despite the completely different numerical approaches used to model dust coagulation. The largest differences arise in the upper atmosphere, where \mcdust{} suffers from low mass resolution and \tripod{} fails to reproduce the exact shape of size distributions that deviate from a power-law. 

\end{abstract}

\begin{keywords}
protoplanetary discs -- planets and satellites: formation -- planets and satellites: general
\end{keywords}




\section{Introduction}
Protoplanetary discs are predominantly composed of gas, with dust contributing at most a few percent to the total mass. Despite its relatively low abundance, dust plays a crucial role in disc evolution, and it is from dust that pebbles and planetesimals - the building material of terrestrial planets and giant planet cores - are formed. Models of dust evolution should therefore be an integral part of any planet formation simulation. Additionally, dust evolution models are indispensable for the interpretation of disc observations, as dust dominates the opacity, which sets the disc’s thermal structure and observational appearance. In this study, we compare three 1D radial and three 2D radial-vertical open-source dust evolution codes to establish whether the choice of code significantly impacts disc evolution, planet formation simulations, and the interpretation of disc observations. Furthermore, we perform a parameter study to determine how each of the key parameters governing dust evolution affects disc evolution and planet formation.

Dust evolution in protoplanetary discs encompasses two key processes: dust coagulation and dust transport. The motions of dust grains in protoplanetary discs are determined by a combination of gas drag (the grains' reaction to the gas accretion flow), radial drift \citep{Weidenschilling1977}, vertical settling \citep{Dubrulle1995}, Brownian motion \citep{Brauer2008}, and turbulent mixing \citep{OrmelCuzzi2007}. The aforementioned processes all depend on the size of the dust grains, with small grains being more tightly coupled to the gas than large grains. As a result, dust grains in protoplanetary discs experience differential velocities, resulting in collisions between particles. 

Dust collisions can lead to sticking via surface forces, resulting in dust growth. Collisional growth does not continue indefinitely but is impeded by several growth barriers. Experiments have found that collisions occurring at velocities above $\sim 1-10\, \textrm{m\,s}^{-1}$ result in fragmentation \citep{BlumWurm2008, GundlachBlum2015, MusiolikWurm2019, Musiolik2021}. During such events, one or both colliding grains fragment into a power-law size distribution, where most of the mass is in the larger fragments \citep{Guttler2010}. Two additional growth barriers -- the relevance of which is still debated -- are the bouncing barrier (e.g. \citealt{BlumMunch1993, Zsom2010, DominikDullemond2024}) and the charge barrier \citep{Okuzumi2011}. The radial drift of grains also acts as a growth barrier when the drift timescale is shorter than the growth timescale. In this study, we only consider the fragmentation and radial drift barriers.  

Some of the most fundamental disc and material properties that influence dust evolution are the gas and dust mass budget and its distribution throughout the disc (i.e., the density structure), the temperature structure, the level of turbulence (which regulates mass transport and vertical mixing), and the mineralogy and structure of individual grains (e.g., their internal density). The dust-growth timescale is primarily determined by the local dust-to-gas density ratio, which can vary across the disc, and the Keplerian frequency. The drift velocity -- which, together with the growth timescale, sets the rate at which dust is depleted -- depends on the pressure gradient, gas density, grain size and internal density. The maximum grain size achievable in the fragmentation-limited regime is controlled by the gas density and temperature, the amount of turbulence, and grain properties such as internal density. Advances in observational capabilities over the past decade, particularly with the Atacama Large millimetre/submillimetre Array (ALMA), have provided improved constraints on these properties, although large uncertainties remain (see discussion in \citealt{Miotello2023}). 


In Section \ref{sec: dust evolution}, we introduce common approaches to modelling dust evolution and the codes used in this study. Planet formation theory, specifically planetesimal formation and pebble accretion, is introduced in Section \ref{sec: planet formation theory}. Our study of 1D radial dust evolution models is presented in Section \ref{sec: 1D radial simulations}. Section \ref{sec: 2d radial-vertical simulations} covers the 2D radial–vertical models. Finally, we discuss the limitations of our study in Section \ref{sec: discussion} and summarize the main findings in Section \ref{sec: conclusion}.


\section{Models of dust evolution}\label{sec: dust evolution}
When modelling collisional dust growth, it is unfeasible to simulate every single dust grain. As an example, to simulate the evolution of dust in the protosolar disc starting from a population of micron-sized grains with an internal density of $1\, \textrm{g\,cm}^{-3}$, one would need to simulate $10^{42}$ grains. To overcome this problem, various techniques have been developed. One approach is to use the Monte Carlo method to simulate the evolution of a much smaller number of representative particles, where each representative particle represents a large number of physically identical particles (e.g. \citealt{Ormel2007,Drazkowska2013,Eriksson2020,Vaikundaraman2025}). The advantages of this method are that the trajectories and histories of individual particles are tracked, and it is straightforward to add additional particle properties at minimal additional computational cost. The disadvantage is that a large number of representative particles is needed to resolve the size distribution over the full disc, making the method computationally expensive. 

A second, less computationally expensive method is to simulate the evolution of a particle size distribution instead of individual particles. Numerically, this is done by solving a discretized form of the Smoluchowski coagulation equation, where the distribution is represented by $N$ different particle species \citep{LombartLaibe2021,StammlerBirnstiel2022,Robinson2024}. The main drawback of this method is that it is difficult to track the histories of particles. Although it is less computationally expensive than the Monte Carlo method, it is still not feasible to perform many large-scale 2D/3D simulations, or big 1D parameter studies with tens of thousands of simulations. For this reason, simpler semi-analytic models have been developed, in which the full particle distribution is not evolved \citep{Birnstiel_2012,Drazkowska2021,Pfeil2024}.

In this paper, we consider the following 1D radial dust evolution codes: \texttt{DustPy} \citep{StammlerBirnstiel2022}, \tripod{} \citep{Pfeil2024,tripodpy2025}, and \texttt{two-pop-py} \citep{Birnstiel_2012}. These codes were selected because they are open-source, widely used in the community, and computationally efficient enough to allow for a relatively large set of simulations to be performed and compared. In addition, we carry out a smaller comparison of 2D radial-vertical dust evolution codes using: \texttt{cuDisc} \citep{Robinson2024}, \tripod{}, and \texttt{mcdust} \citep{Vaikundaraman2025_mcdust}. A GitHub directory containing scripts for initializing and running all simulations and post-processing presented in this paper is available at \url{https://github.com/astrolinn/dustComparison.git.}


\subsection[DustPy]{\texttt{DustPy}}\label{subsec:dustpy}
\texttt{DustPy} is a 1D dust evolution code that simulates the evolution of the dust mass distribution in a disc, accounting for dust transport, collisional growth, and viscous evolution of the gas disc. Dust growth is calculated by solving the Smoluchowski equation, and dust transport is calculated by solving the advection-diffusion equation. The dust mass distribution is discretized by $N$ mass bins, and a minimum of 7 mass bins per mass decade should be used \citep{Ohtsuki1990,Drazkowska2014}. The evolution of the gas disc is calculated by solving the viscous advection-diffusion equation. 

The default version of \texttt{DustPy} distinguishes between two collisional outcomes: perfect sticking and fragmentation. In the case of fragmentation, the code further distinguishes between full fragmentation and erosion. Full fragmentation occurs when the colliding particles have similar masses, resulting in the complete break-up of both particles into a fragment distribution that follows a power law with exponent $-11/6$ \citep{Dohnanyi1969}. Erosion occurs when the particle mass ratio is larger than 10, in which case the smaller particle fully fragments into a fragment distribution while some mass is chipped off the larger particle. 

The coupling between particles and gas is governed by the Stokes number, $\St$, with $\St=1$ corresponding to the maximum radial drift velocity. \texttt{DustPy}, \tripod{}, and \texttt{two-pop-py} all consider two aerodynamic drag regimes, the Epstein regime and the Stokes I regime. The corresponding equations for the $\St$ of a particle of size $a$ are
\begin{equation}\label{eq: Stokes number}
\St =
\begin{cases}
\frac{\pi}{2} \frac{a \rho_{\bullet}}{\Sigma_{\rm gas}}, 
& \quad a < \frac{9}{4} \lambda_{\mathrm{mfp}}\,\text{(Epstein)} \\[1em]
\frac{2\pi}{9} \frac{a^2 \rho_{\bullet}}{\lambda_{\mathrm{mfp}} \Sigma_{\rm gas}}, 
& \quad \text{otherwise}\,\text{(Stokes I)}
\end{cases}
\end{equation}
where $\rho_{\bullet}$ is the internal density of the dust grain, $\Sigma_{\rm gas}$ is the gas surface density and $\lambda_{\mathrm{mfp}}$ is the mean free path of the gas. The mean free path is calculated as
\begin{equation}
    \lambda_{\mathrm{mfp}} = \mu m_p / (\sqrt{2}\rho_{\rm gas} \sigma_{\rm H}),
\end{equation}
where $\mu$ is the mean molecular weight, $m_{\rm p}$ is the proton mass, $\rho_{\rm gas}$ is the midplane gas density and $\sigma_{\rm H}$ is the collisional cross section of the Hydrogen atom. 


\subsection[TriPoD]{\tripod{}}
\label{subsec: TriPoDIntro}
Dust size distributions in coagulation-fragmentation equilibrium can be approximated as truncated power laws of the particle size $n(a)\propto a^q$ \citep{Birnstiel2011}. \tripod{} is a semi-analytic dust coagulation model utilizing this power-law prescription to model dust growth and fragmentation with only two dust fluids (large and small particles, $\Sigma_0$, $\Sigma_1$) and a maximum particle size $a_\mathrm{max}$ \citep{Pfeil2024}. A polydisperse size distribution can be reconstructed from the three variables by calculating the power-law exponent 
\begin{equation}
    q=2\frac{\log(\Sigma_1/\Sigma_0)}{\log(a_\mathrm{max}/a_\mathrm{min})} - 4,
\end{equation}
which makes a detailed comparison to codes like \texttt{DustPy}, \texttt{cuDisc}, and \texttt{mcdust}, possible.
The collisional velocities and density ratios of the two dust fluids determine how mass is exchanged between the large and small populations. 
The maximum particle size is treated as a passive scalar, following the larger dust fluid. It is evolved based on the monodisperse growth rate \citep{Kornet2001}, which is based on the total dust density and the collision velocities between large particles. 
Since these mass transfer rates and the size evolution are expressed as simple analytical collision rates, they were manually calibrated to approximate the actual numerical solutions to the Smoluchowski equation.
\cite{Pfeil2024} also conducted comparisons between \tripod{} and \texttt{DustPy} simulations for varying fragmentation velocities and disc models with substructure and found a good agreement between the two methods.
\tripod{}'s accuracy is, however, limited whenever the power-law prescription cannot approximate the size distribution. Multimodal grain size distributions, and the detailed substructures within the size distributions can thus not be well reproduced with this method.

In this study, we use two different versions of \tripod{} for the 1D and 2D setups. We use the open-source \texttt{Python} implementation \texttt{TriPoDPy} \citep{tripodpy2025} for the radial simulations, as it shares most of the physics and setup with \dustpy, making comparisons of the simulations easier. For the 2D simulations, we use the three-dimensional version of \tripod{} (\texttt{Athena++} based) that was not included in the original \cite{Pfeil2024} publication. 
For this, minimal modifications to the code had to be made to run radial-vertical setups. All column densities are exchanged for volume densities. Factors including the dust scale heights are dropped. 
The relative particle settling velocities are exchanged for the local prescription
\begin{equation}
    \Delta v_{\mathrm{set},0,1} = \max(c_\mathrm{s}, (\mathrm{St}_1 - \mathrm{St}_0)\Omega_\mathrm{K}z).
\end{equation}
We have conducted a slight recalibration of the method to gain a better agreement with \cudisc{}. For this, the diffusion of the maximum particle size is now based on the gradient in the mass-averaged particle size instead of the gradient in maximum grain size itself. This takes into account the vertically varying power-law exponents of the size distributions.
Furthermore, the shrinkage parameter, which had been calibrated to a factor of $f_\mathrm{crit}=0.425$ in vertically integrated setups is now set to $f_\mathrm{crit}=0.495$, which results in a better fit with \cudisc{}, especially in the low-diffusivity case (see \autoref{sec: 2d radial-vertical simulations}).


\subsection[two-pop-py]{\texttt{two-pop-py}}
The 1D dust evolution code \texttt{two-pop-py} avoids the treatment of a polydisperse size distribution completely by assuming fixed mass ratios between micron-sized grains and grown grains. ``Fudge factors'' are employed that determine the fraction of large dust and small dust, based on the evolutionary regime (growing, drift-limited, or fragmentation limited).
The fraction of large dust determines the magnitude of the radial dust flux under the assumption that small grains do not participate in radial drift and only follow the gas.
The fudge factors are fitted to reproduce a standard dust coagulation model.
Particle growth is realized via the monodisperse growth rate and stopped at the analytically calculated drift, fragmentation, and drift-fragmentation limits.
The drift and fragmentation limits are given by
\begin{equation}\label{eq: fragmentation barrier}
a_{\mathrm{frag}} = f_\mathrm{f}\frac{2}{3 \pi} \frac{\Sigma_{\rm gas}}{\rho_{\bullet} \alpha_{\rm t}} \frac{v_{\mathrm{frag}}^2}{c_s^2},
\end{equation}
and
\begin{equation}\label{eq: drift barrier}
a_{\mathrm{drift}} = f_\mathrm{d}\frac{2}{\pi}\frac{\Sigma_{\mathrm{dust}}}{\rho_{\mathrm{\bullet}}} \left( \frac{H_{\mathrm{dust}}}{r} \right)^{-2} 
\left| \frac{\partial \log P}{\partial \log r} \right|^{-1}.
\end{equation}
In the above equations $f_\mathrm{f}$ and $f_\mathrm{d}$ are fudge factors, $\alpha_{\rm t}$ is the turbulent parameter, $v_{\rm frag}$ is the fragmentation velocity, $c_{\rm s}$ is the sound speed, $\Sigma_{\rm dust}$ is the dust surface density, $H_{\rm dust}$ is the dust scale height, $r$ is the semimajor axis and $P$ is the gas pressure. The dust scale height is given by
\begin{equation}
    H_{\rm dust} = H \sqrt{\frac{\alpha_{\rm t}}{\St + \alpha_{\rm t}}},
\end{equation}
where $H$ is the gas scale height.
\texttt{two-pop-py} offers the fastest possible method to mimic the effects of dust growth in simulations of protoplanetary discs and was shown to reproduce the growth pattern and the radial fluxes of a small number of simulations with a full Smoluchowski solver.


\subsection[cuDisc]{\texttt{cuDisc}}
\cite{Robinson2024} introduced \cudisc{}, a two-dimensional (radial–vertical) dust evolution code designed to run on GPUs. \cudisc{} solves the Smoluchowski equation for dust growth on a discrete mass grid, similar to \dustpy{}, and computes dust dynamics using a second-order finite-volume Godunov solver. The evolution of the gas disc is solved using the viscous evolution equation, and hydrostatic equilibrium is assumed in the vertical direction. The code also includes a flux-limited diffusion solver and a ray tracing algorithm to derive the disc's temperature structure. For this, methods for calculating dust absorption and scattering opacities are provided. The dust size distribution evolves due to coagulation and fragmentation, and particle relative velocities are computed as the sum of contributions from Brownian motion, gas turbulence, and laminar dust motion. 


\subsection[mcdust]{\texttt{mcdust}}
\texttt{mcdust} is a parallel two-dimensional (radial–vertical) Monte Carlo dust evolution code that employs a representative particle approach to model the growth and transport of a limited number of particles \citep{Vaikundaraman2025_mcdust}. The default model includes three possible collision outcomes: perfect sticking, fragmentation, and erosion. Fragmentation and erosion are modelled similarly to \texttt{Dustpy} as described in Sec. \ref{subsec:dustpy}. \texttt{mcdust} assumes a static gas disc with fixed density and temperature profiles in its default version. The Lagrangian particles are grouped into grid cells to determine local gas properties and perform collisions. To maintain a uniform number of representative particles per cell, the code utilizes an adaptive grid. Collisional probabilities are computed for every particle pair within each cell based on their particle sizes and relative velocities, and a random number determines which pair undergoes a collision.

\texttt{mcdust} solves dust dynamics in the radial and vertical direction, including turbulent diffusion. Transport by gas advection and radial drift is modelled in the radial direction, with settling due to the gravity from the central star modelled in the vertical direction. The effect of turbulence on the particles is modelled as random kicks in the same way as \cite{Ciesla2010} and \cite{Zsom2011}. Turbulent diffusion in both radial and vertical directions is taken into account for dust transport. We refer the reader to \cite{Drazkowska2013} for the details of the physics of the code. 


\section{Planet formation theory}\label{sec: planet formation theory}


\subsection{Planetesimal formation}\label{subsec: planetesimal formation theory}
The continued growth from pebbles to planetesimals via direct growth is hindered by the radial drift and fragmentation barriers. Some studies have suggested that these barriers could be bypassed if the dust grains are highly porous; however, these results remain controversial (see discussion in \citealt{Drazkowska2023}). Currently, the favored mechanism for planetesimal formation is the gravitational collapse of pebble clumps formed by the streaming instability (SI, \citealt{YoudinGoodman2005, JohansenYoudin2007}). The SI produces planetesimal populations that are consistent with several observed properties of Solar System minor bodies, such as the architecture of trans-Neptunian binaries \citep{Nesvorny2019} and the size distribution of the cold classical Kuiper belt \citep{Kavelaars2021}. 

The parameter space conducive to strong particle concentration via the SI remains an active area of research. Studies by \citet{YoudinGoodman2005} and \cite{JohansenYoudin2007} showed that the growth rate of the SI in the linear regime and the clumping of the SI in the nonlinear regime are stronger when $\epsilon \gtrsim 1$, where $\epsilon$ is the midplane dust-to-gas density ratio. Based on this we adopt $\epsilon \geq 1$ as our first SI criteria, and refer to it as YG05. The clumping criteria for the nonlinear SI have been studied extensively using 2D stratified simulations \citep{Carrera2015, Yang2017, LiYoudin2021, Lim2025}, and lately also using 3D stratified simulations \citep{Lim2025_3D}. These studies present the clumping threshold $Z_{\rm crit}$ as a function of the Stokes number and are valid for mono-disperse size distributions in the absence of external turbulence. The most comprehensive of these criteria is that of \citet{Lim2025}
\begin{equation}\label{eq: L25}
    \log(Z_{\rm crit}) \approx 0.10 (\log \St)^2
    + 0.07 \log \St - 2.36
\end{equation}
(hereafter referred to as L25), which was tested over the range $\St \in [10^{-3},1]$ and derived using a global radial pressure gradient of $0.05$. 

The above criterion is valid only in the absence of external turbulence. \citet{Lim2024} performed a large suite of 3D simulations of the SI under externally driven (forced) turbulence and obtained the criterion
\begin{equation}
\begin{split}
    \log Z_{\rm crit}(\St, \alpha_{\rm t}) 
    &= 0.15 (\log \alpha_{\rm t})^2 - 0.24 \log \St \log \alpha_{\rm t} \\
    &- 1.48 \log \St + 1.18 \log \alpha_{\rm t} 
\end{split}
\end{equation}
(hereafter referred to as L24). This criterion was tested over the range $\alpha_{\rm t} \in [10^{-4},10^{-3}]$ and $\St \in [10^{-2},10^{-1}]$ and is derived using the same global radial pressure gradient as L25. \citet{Lim2024} found that externally forced turbulence weakens clumping by SI. However, turbulence driven by the magnetorotational instability (MRI) has been shown to concentrate particles \citep{Johansen2007,Yang2018,XuBai2022}, and in this scenario clumping occurs at significantly lower $Z_{\rm crit}$ than in forced-turbulence simulations \citep{Eriksson2026}.

The previously mentioned SI criteria are derived from simulations considering particles of a single Stokes number (mono-disperse SI). However, discs contain a distribution of particle sizes. The poly-disperse SI has been explored through both linear \citep{Krapp2019,ZhuYang2021,YangZhu2021} and nonlinear \citep{BaiStone2010,Schaffer2021} simulations, yielding mixed results. Some studies report that the SI is significantly suppressed in the poly-disperse case compared to the mono-disperse case \citep{Krapp2019}, while others find that a particle size distribution does not necessarily lead to overall unfavorable clumping conditions \citep{Schaffer2021}. Notably, all of these studies assume a static size distribution. However, recent work suggests that dust growth and the SI can promote each other \citep{Ho2024,Carrera2025,Tominaga2025,Vallucci-Goy2026,Pierens2026}, potentially leading to more favorable conditions for planetesimal formation. 

Since a fit to the clumping boundary for the poly-disperse SI is not available, we adopt the SI criteria derived for mono-disperse size distributions in this work. To apply these criteria in combination with a code like \texttt{DustPy}, which evolves a full size distribution, a representative $\St$ must be selected. Two common choices are the density-weighted average $\St$ and the $\St$ corresponding to the peak of the size distribution. Depending on the shape of the distribution, the density-weighted average can either coincide with the peak or be significantly smaller. Consequently, the choice of representative $\St$ can have a significant effect on the resulting $Z_{\rm crit}$.


\subsection{Pebble accretion}\label{subsec: pebble accretion theory}
Once a planetesimal has formed, there are two main pathways for further solid growth: planetesimal accretion \citep{Pollack1996} and pebble accretion \citep{JohansenLacerda2010,OrmelKlahr2010,Lambrechts2012}. While pebble accretion can account for the formation of a wide range of planetary types both in the Solar System and beyond (see discussion in \citealt{Drazkowska2023} and \citealt{Lyra2023}), planetesimal accretion alone cannot explain the origin of wide-orbit giants such as Uranus and Neptune \citep{JohansenBitsch2019}. The efficiency of pebble accretion stems from the dissipation of energy by gas drag as pebbles approach the planet, which allows pebbles to be accreted from a much larger effective radius than planetesimals. The accretion rate of pebbles depends sensitively on the pebble flux, size, and vertical distribution, all of which are governed by dust evolution.

The mono-disperse pebble accretion rate is given by
\begin{equation}
    \dot{M}_{\rm pebb} = \pi R_{\rm acc}^2 \rho_{\rm dust} \delta v e^{-\xi}[I_0(\xi)+I_1(\xi)],
\end{equation}
where
\begin{equation}
    \xi \equiv \left(\frac{R_{\rm acc}}{2H_{\rm dust}}\right)^2
\end{equation}
\citep{Lyra2023}. In the above equations $R_{\rm acc}$ is the accretion radius, $\rho_{\rm dust}$ is the midplane dust density, $\delta v$ is the approach velocity, and $I_0(\xi)$ and $I_1(\xi)$ are the modified Bessel functions of the first kind. The approach velocity is given by
\begin{equation}
    \delta v = \Delta v + \Omega_{\rm K}R_{\rm acc},
\end{equation}
where
\begin{equation}
    \Delta v = -\frac{c_{\rm s}}{2}\frac{H}{r}\frac{d\ln P}{d\ln r}
\end{equation}
is the sub-Keplerian velocity reduction, $\Omega_{\rm K}$ is the Keplerian angular velocity, $P=\Sigma_{\rm gas}T/H$ is the gas pressure, and $T$ is the midplane temperature. The accretion radius is given by
\begin{equation}
R_{\rm acc} = 
\begin{cases}
\left(\frac{4\tau_\mathrm{f}}{t_{\rm B}}\right)^{1/2}R_{\rm B} \exp\left[-\chi\left(\frac{\tau_\mathrm{f}}{t_\mathrm{p}}\right)^{\gamma}\right]
& \quad \text{if} \;\, M_{\rm p}<M_{\rm HB} \;\, (\text{Bondi}), \\[1em]
\left(\frac{\St}{0.1}\right)^{1/3}R_{\rm H} \exp\left[-\chi\left(\frac{\tau_\mathrm{f}}{t_\mathrm{p}}\right)^{\gamma}\right]
& \quad \text{if} \;\, M_{\rm p}\geq M_{\rm HB} \;\, (\text{Hill}),
\end{cases}
\end{equation}
where $\tau_\mathrm{f}=\St/\Omega_{\rm K}$ is the friction time, $t_{\rm B}=R_{\rm B}/\Delta v$ is the Bondi time, $R_{\rm B}=GM_{\rm p}/\Delta v^2$ is the Bondi radius, $G$ is the gravitational constant, $M_{\rm p}$ is the planetary mass, $\chi=0.4$, $t_\mathrm{p}=GM_{\rm p}/(\Delta v + \Omega_{\rm K} R_{\rm H})^3$ is the characteristic passing timescale, $R_{\rm H}=(GM_{\rm p}/(3\Omega_{\rm K}^2))^{1/3}$ is the Hill radius, $\gamma=0.65$, $M_{\rm HB}=M_{\rm t}/(8\St)$ is the transition mass between Bondi and Hill accretion \citep{Ormel2017}, and $M_{\rm t}=\Delta v^3/(G\Omega_{\rm K})$. 

We obtain the poly-disperse accretion rate by calculating the mono-disperse accretion rate for all particle sizes in the distribution and summing up the results. To prevent numerical issues, we truncate the size distribution once the cumulative dust mass fraction reaches 99.9\% (thus removing the high-size tail) for \tripod{} and \texttt{DustPy}, and use the standard 2D limit for pebble accretion (Eq. 30 of \citealt{Lyra2023}) when $\xi>10$. We do not consider the focusing regime for pebble accretion.


\section{1D radial simulations}\label{sec: 1D radial simulations}


\subsection{Simulation set-up}\label{subsec: 1D simulation set-up}
\texttt{DustPy}, \tripod{}, and \texttt{two-pop-py} all simulate the evolution of gas and dust in a 1D viscous protoplanetary disc. We pick the initial gas surface density profile to be
\begin{equation}
    \Sigma_{\rm gas, 0} = \frac{\dot{M}_0}{3\pi\nu_{\rm c}(r/r_{\rm c})^{\gamma}}\exp\left[-\left(\frac{r}{r_{\rm c}}\right)^{2-\gamma}\right],
\end{equation}
where $\dot{M}_0$ is the initial disc accretion rate, $\nu_{\rm c}$ is the kinematic viscosity at the exponential cut-off radius $r_{\rm c}$, and $\gamma$ is the radial viscosity gradient \citep{Lynden-Bell+1974}. The initial dust surface density is taken to be $\Sigma_{\rm dust}=Z\Sigma_{\rm gas}$, where $Z$ is the initial dust-to-gas surface density ratio. 

The viscosity is calculated as $\nu = \alpha \Omega_{\rm K}H^2$, where $\alpha$ is the viscosity parameter \citep{ShakuraSunyaev1973}, and $H=c_{\rm s}/\Omega_{\rm K}$. The sound speed is calculated as $c_{\rm s} = \sqrt{k_{\rm B}T/(\mu m_{\rm p})}$, where $k_{\rm B}$ is the Bolzmann constant. We adopt $\alpha=5\times 10^{-3}$ and $\mu=2.34$ throughout this work. 
We consider the temperature profile of a passively irradiated disc with an irradiation angle of $0.05$,

\begin{equation}
    T = \sqrt[4]{\frac{1}{2}\frac{0.05L_{\ast}}{4\pi r^2 \sigma_{\rm SB}}},
\end{equation}
\begin{equation}
    L_{\ast} = 4\pi R_{\ast}^2 \sigma_{\rm SB} T_{\ast}^4,
\end{equation}

\noindent where $L_{\ast}$, $R_{\ast}$, and $T_{\ast}$ are the luminosity, radius, and effective temperature of the central star, respectively, and $\sigma_{\rm SB}$ is the Stefan-Boltzmann constant \citep{ChiangGoldreich1997}. 

Our semimajor axis grid stretches from $1-1000\, \textrm{au}$, with 150 logarithmically spaced grid cells in the range $1-200\, \textrm{au}$ and 16 linearly spaced grid cells in the range $200-1000\, \textrm{au}$. We adopt a large radial extent in order to prevent numerical issues near the outer boundary. The dust-grid in \texttt{DustPy} and \tripod{} has a minimum size $a_{\rm min}=0.5\, \mu\textrm{m}$ in the nominal set-up. We caution that, for some parameter combinations, reducing $a_{\rm min}$ in \texttt{DustPy} is necessary to prevent an unphysical pile-up of mass at the smallest particle sizes. Because this substantially increases the computational cost, we retain the nominal $a_{\rm min}$ in the main simulations; additional simulations with smaller $a_{\rm min}$ confirm that the main results remain unchanged. We simulate the evolution of the disc for $3\, \textrm{Myr}$, which we take to be the disc's lifetime. We note that the default versions of the dust evolution codes do not model disc dispersal - which could occur via e.g. photoevaporation. Our simulations might thus still contain significant amounts of gas and dust at $3\, \textrm{Myr}$.  


\subsection{Parameter choices}\label{subsec: 1D parameter choices}

\begin{table}
\centering
\caption{Parameters used in the 1D simulations. Left column: List of parameters that are varied in the parameter study. Middle column: Parameter values of the nominal simulation. Right column: Range of values considered in the parameter study.}
\label{table: nominal}
\begin{tabular}{lll}
\hline \hline
 & Nominal simulation & Parameter study                                \\ \hline
$M_{\ast}$               & $1\, \textrm{M}_{\odot}$ & $0.05-1.3\, \textrm{M}_{\odot}$ ($R_{\ast}$, $T_{\ast}$, $\dot{M}_0$ also varied)                               \\
$R_{\ast}$               & $3.1\, \textrm{R}_{\odot}$ & from \citet{Baraffe2015}                           \\
$T_{\ast}$               & $4397\, \textrm{K}$ & from \citet{Baraffe2015}                                   \\
$\dot{M}_0$         & $4\times 10^{-8}\, \textrm{M}_{\odot}\,\textrm{yr}^{-1}$ & $8\times 10^{-10}-8\times 10^{-8}\, \textrm{M}_{\odot}\,\textrm{yr}^{-1}$ \\
$r_{\rm c}$         & $50\, \textrm{au}$ & $10-250\, \textrm{au}$                                 \\
$Z$                 & $10^{-2}$ & $10^{-4}-10^{-1}$                                                 \\
$\alpha_{\rm t}$ & $10^{-4}$ & $10^{-5}-10^{-2}$                                             \\
$v_{\rm frag}$      & $1\, \textrm{m}\,\textrm{s}^{-1}$ & $0.1-25\, \textrm{m}\,\textrm{s}^{-1}$                     \\
$\rho_{\bullet}$    & $1\, \textrm{g}\,\textrm{cm}^{-3}$ & $0.1-10\, \textrm{g}\,\textrm{cm}^{-3}$                      \\
$a_0$               & $1\, \mu\textrm{m}$ & $0.1\, \mu\textrm{m}$                                  \\
adp                 & False & True                                                   \\
Nd                  & $7$ & $10$                                                                                          \\ \hline
\end{tabular}
\end{table}

The parameter values for the nominal simulation are presented in \autoref{table: nominal}, along with the range of values considered in the parameter study. The complete setup for the nominal simulation can be found in the associated GitHub repository. In the parameter study, we vary one parameter at a time, unless otherwise specified, in order to study the effect of that parameter on planet formation and code agreement. We vary the stellar mass $M_{\ast}$ between $0.05-1.3\, \textrm{M}_{\oplus}$ and adjust $\dot{M}_0$ accordingly to maintain an initial disc-to-star mass ratio of 5\%. The corresponding stellar radii and effective temperatures are adopted from \citet{Baraffe2015}. The range of $\dot{M}_0$ values is chosen to yield initial disc-to-star mass ratios between 0.1-10\%, and lies within the range of observed accretion rates \citep{Manara2023}. The exponential cut-off radius is varied between $10-250\, \textrm{au}$, corresponding to initial disc-to-star mass ratios of 0.01-25\%. Increasing $r_{\rm c}$ results in a more extended disc, with higher surface densities at large semimajor axes, while the inner disc remains largely unaffected.

We consider a broad range of $Z$, including very dust-poor systems that are relevant for studies of planet formation at high redshift \citep{Eriksson2025}. The turbulent parameter $\alpha_{\rm t}$ governs the vertical settling, radial diffusion and turbulent mixing of dust particles. An increase in $\alpha_{\rm t}$ results in a higher dust scale height and larger relative velocities between particles. We vary $\alpha_{\rm t}$ between $10^{-5}-10^{-2}$, which covers the range reported from protoplanetary disc observations (e.g. \citealt{Villenave+22, Pinte+23}). Collision experiments typically yield fragmentation velocities between $1-10\, \textrm{m\,s}^{-1}$ \citep{BlumWurm2008, GundlachBlum2015, MusiolikWurm2019, Musiolik2021}; however, there are significant uncertainties in some of the underlying parameters, such as the surface free energy \citep{BlumWurm2000, Kimura2015, Morrissey2025}. Therefore, we explore a broader range of $v_{\rm frag}$ between $0.1-25\, \textrm{m\,s}^{-1}$. In our study, as in most others, $v_{\rm frag}$ is treated as a constant. Whether this assumption is valid, or whether $v_{\rm frag}$ instead varies with properties such as disc temperature, grain porosity, and grain composition, is still under debate \citep{GundlachBlum2015, MusiolikWurm2019, Musiolik2021}.

Experiments show that porosity impacts the outcome of dust collisions, with porous particles typically growing more efficiently than compact particles \citep{Guttler2010}. In reality, collisions between porous grains likely result in some degree of compaction - an effect that is not accounted for in any of the three dust evolution codes (see \citealt{Michoulier2024} for an example of a dust evolution study including compaction). Although the effect of porosity is not studied in detail in this work, we do vary the internal dust density, $\rho_{\bullet}$. Finally, we study the effect of decreasing the initial dust size $a_0$, increasing the number of mass bins per mass decade (Nd, relevant only for \texttt{DustPy}), and whether or not drifting particles are allowed at $t=0$ (adp). When adp=True, dust of size $a_0$ is initialized throughout the disc. In the outer regions of the disc, where the Stokes numbers are large, even small grains are in the radial drift regime, and initializing particles that are already drifting may not be physical. Therefore, we primarily perform simulations where initially drifting particles are removed (adp=False), and we have implemented this functionality in our version of \texttt{two-pop-py}. 


\subsection{The sizes of particles in the disc}\label{subsec: 1D particle size}

\begin{figure*}
    \centering
    \includegraphics[width=0.87\textwidth, trim=0 0.2cm 0 0.2cm, clip]{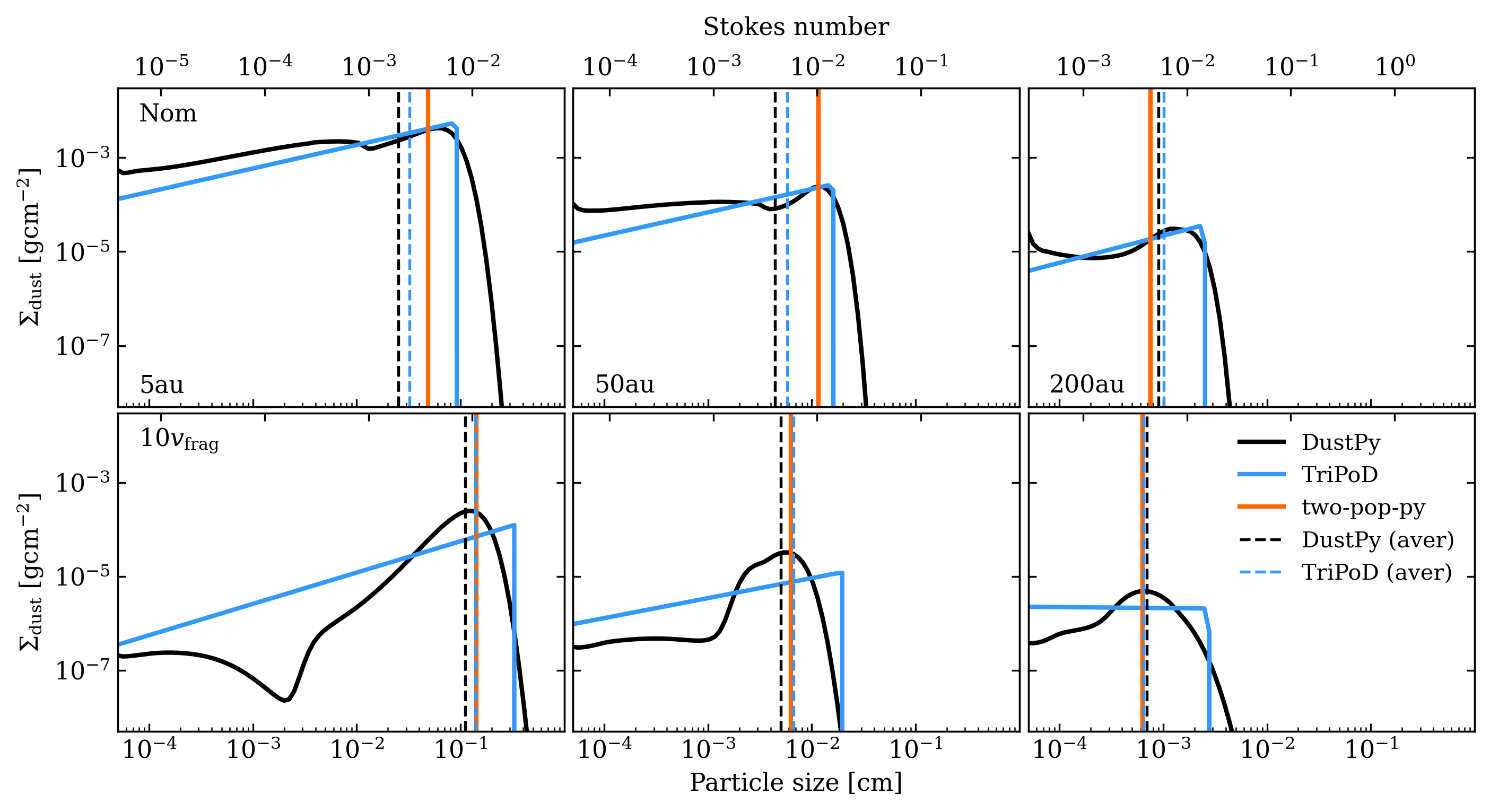}
    \caption{Comparison of the particle size distribution at $3\, \textrm{Myr}$ from the three different 1D codes, at three different semimajor axis and for two different simulations. The bottom axis shows the particle size and the top axis shows the corresponding Stokes number. The dashed lines show the density-weighted average for \tripod{} and \texttt{DustPy}. }
    \label{fig: particle distribution 2 sim}
\end{figure*}
\begin{figure*}
    \centering
    \includegraphics[width=1\textwidth, trim=0 0.2cm 0 0.2cm, clip]{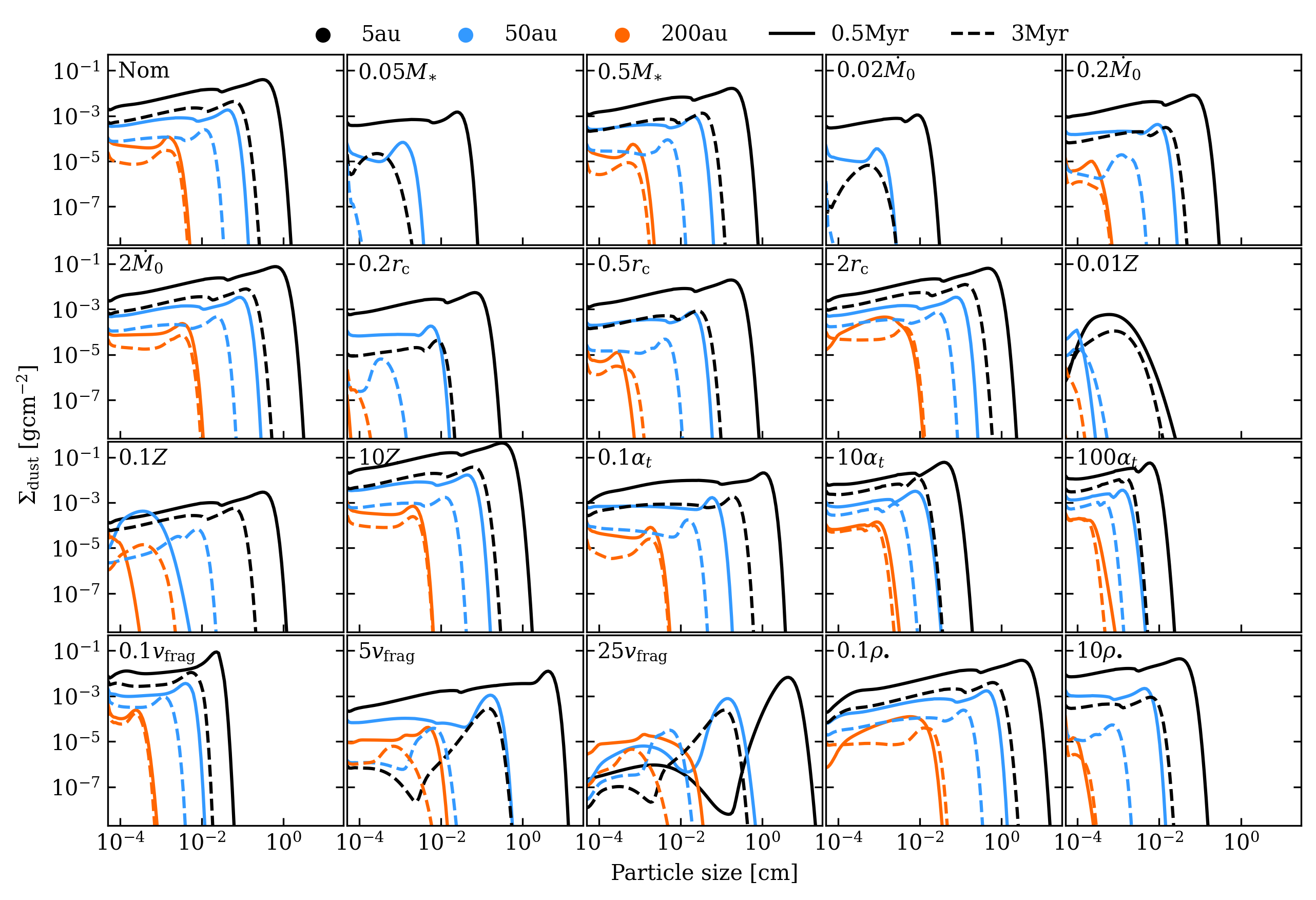}
    \caption{Particle size distribution for a subset of simulations from the parameter study, produced using \texttt{DustPy} and shown for three different semimajor axes and two different times. The simulation name indicated in the top left corner indicate how the varied parameter compares with the Nominal simulation; e.g., simulation $0.01Z$ has $Z = 0.01 \times Z_{\rm Nom}$. }
    \label{fig: size dist param}
\end{figure*}

Dust in the protoplanetary disc is inherited from the host molecular cloud, and the initial dust sizes are typically assumed to be comparable to those found in the interstellar medium (ISM, $~0.1\, \mu \textrm{m}$). As dust grains collide and grow, the size distribution gradually evolves away from the ISM distribution. The growth rate for equal-sized particles in turbulent motion is $\sim 1/(\Omega_{\rm K} Z)$ \citep{Birnstiel_2012}, so growth proceeds fastest in the inner disc and becomes progressively slower with increasing semimajor axis, assuming a constant dust-to-gas ratio $Z$ throughout the disc. The maximum particle size is set by the lower of the radial drift or fragmentation barriers: the fragmentation barrier typically limits growth in the inner disc, while the radial drift barrier dominates in the outermost regions.

The particle size distributions from two of the simulations in the parameter study are shown in \autoref{fig: particle distribution 2 sim}. While \texttt{DustPy} evolves a full particle size distribution (black solid line), \tripod{} and \texttt{two-pop-py} employ simpler and more computationally efficient dust evolution models. For \tripod{}, a smooth polydisperse size distribution can be reconstructed using the available data (blue solid line). In contrast, \texttt{two-pop-py} tracks only the largest particle size, which is also the only output available (orange line). In the nominal simulation (Nom), although the size distribution is not smooth, it is reasonably well represented by a power law. The sharp density increase toward the smallest sizes at $200\, \textrm{au}$ stems from our choice of $a_{\rm min} = 0.5\, \mu \textrm{m}$, and a smaller $a_{\rm min}$, as in our simulation with $a_0=10^{-5}$, would instead show a steep decline in density toward the smallest sizes caused by Brownian motion. The density-weighted average particle sizes (dashed lines) from \texttt{DustPy} and \tripod{} agree relatively well at all three semimajor axes, and are slightly smaller than the particle size at the peak of the size distribution. 

Focusing on the bottom row of \autoref{fig: particle distribution 2 sim}, in which the fragmentation velocity $v_{\rm frag}$ was increased from $1\, \textrm{m\,s}^{-1}$ to $10\, \textrm{m\,s}^{-1}$, the agreement between codes is noticeably worse. In this simulation, growth is limited by radial drift rather than fragmentation in most regions of the disc, resulting in a bimodal size distribution. Such a size distribution is not well represented by a power law, and \tripod{} predicts larger peak particle sizes than \texttt{DustPy}. The density-weighted average particle sizes do however still agree rather well.  

In \autoref{fig: size dist param}, we show the particle size distribution for a subset of simulations in the parameter study (see \autoref{fig: st dist param} for the corresponding Stokes number distribution). The name of the simulations (shown on the plots) indicate how the varied parameter compares with the Nominal simulation; e.g., simulation $2Z$ has $Z = 2 \times Z_{\rm Nom}$. At $5\, \textrm{au}$, most simulations produce a peaked, non-smooth size distribution where the density increases steadily at small and intermediate sizes, followed by a slight dip and then a sharp rise toward the peak. Beyond the peak, the density declines rapidly with increasing particle size. One notable exception is simulation $0.01Z$, where the growth rate is too slow for either growth barrier to be reached, and a power law size distribution never develops. Another exception occurs when the fragmentation velocity is high (simulations $5v_{\rm frag}$ and $25v_{\rm frag}$), in which case growth in most regions of the disc is limited by radial drift rather than fragmentation, and the size distribution evolves to become bimodal. At $50\, \textrm{au}$ and $200\, \textrm{au}$, where the densities are lower, the size distribution tends to flatten more at intermediate sizes, followed by a sharper rise toward the peak.

Focusing on the effect of varying parameters, an increase in $M_{\ast}$, $\dot{M}_0$, or $r_{\rm c}$ - which corresponds to higher initial gas and dust densities - results in larger particle sizes throughout the disc. This is expected as dust growth is fragmentation limited in most of the disc and $a_{\rm frag} \propto \Sigma_{\rm gas}$. Increasing $Z$ does not lead to a significant change in particle size once a size distribution has been established. Higher turbulence levels result in progressively smaller particle sizes, as increased turbulence leads to higher collision velocities, causing particles to reach the fragmentation barrier at smaller sizes ($a_{\rm frag} \propto \alpha_{\rm t}^{-1}$). Conversely, larger fragmentation velocities allow particles to grow to significantly larger sizes ($a_{\rm frag} \propto v_{\rm frag}^{2}$). Finally, increasing the internal density of dust grains results in smaller particles, since both $a_{\rm frag}$ and $a_{\rm drift}$ scale as $\propto \rho_{\bullet}^{-1}$. 


\subsection{Mass of the dust disc}\label{subsec: 1D disc mass}

\begin{figure*}
    \centering
    \includegraphics[width=0.99\textwidth, trim=0 0.3cm 0 0.2cm, clip]{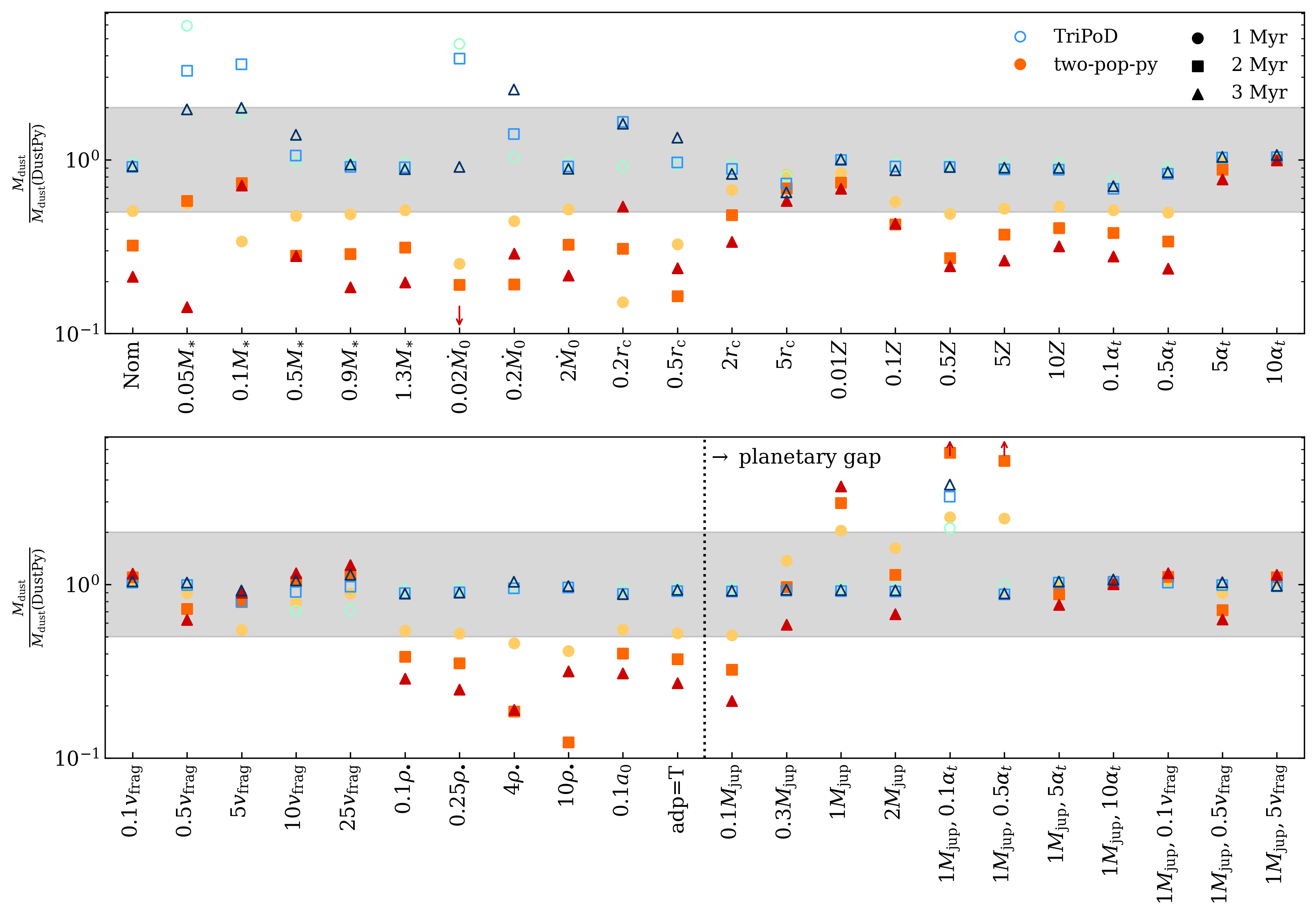}
    \caption{Comparison of the total dust mass - calculated by integrating the dust surface density - at three different times for \texttt{two-pop-py} and \tripod{}, shown relative to \texttt{DustPy}. The gray shaded region indicates where the total dust mass differs by less than a factor of 2. The agreement between \texttt{DustPy} and \tripod{} is generally good, whereas discs simulated with \texttt{two-pop-py} typically deplete their dust mass too quickly.}
    \label{fig: Mdust comparison}
\end{figure*}

\begin{figure*}
    \centering
    \includegraphics[width=0.99\textwidth, trim=0 0.2cm 0 0.2cm, clip]{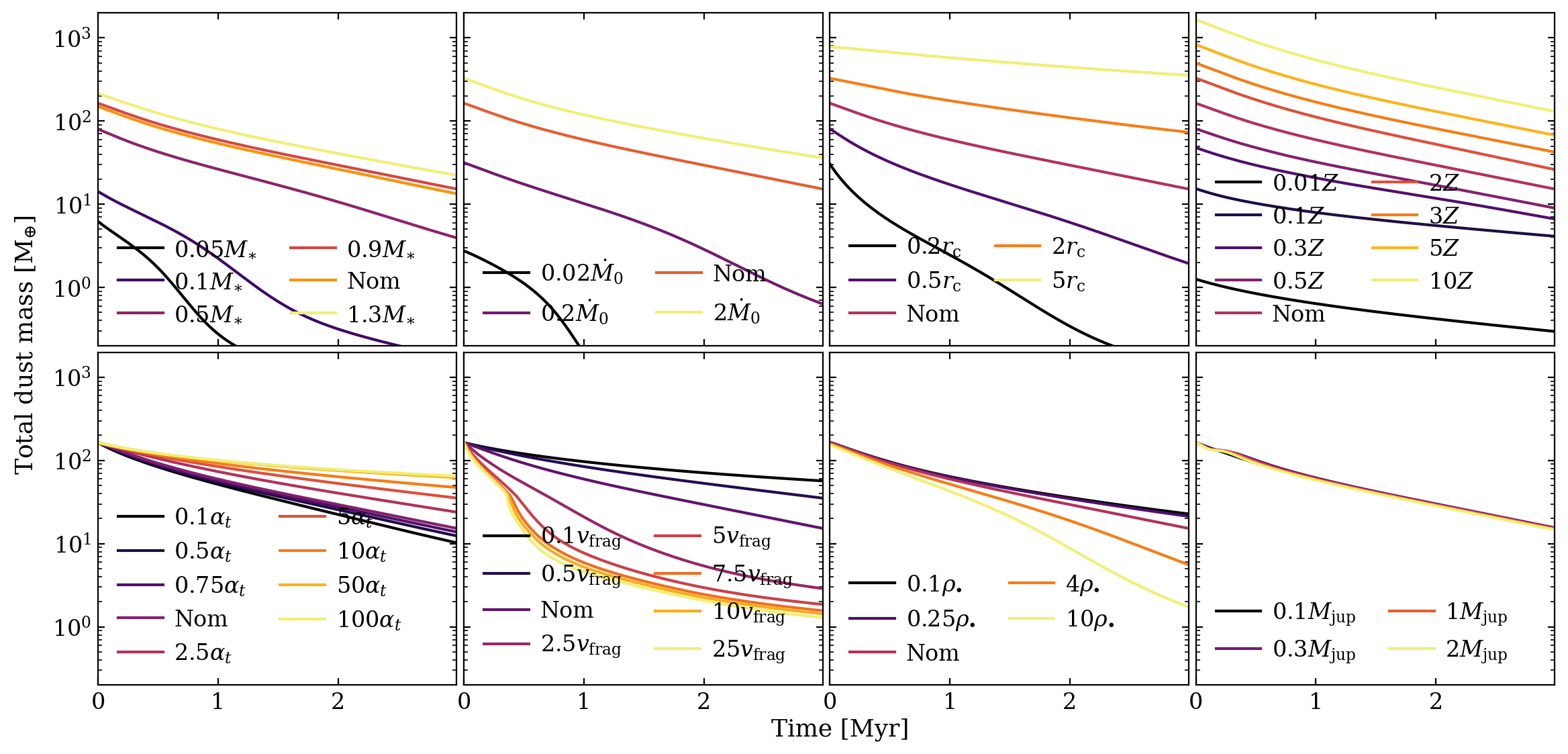}
    \caption{Total dust mass versus time for all simulations in the parameter study, and some simulations including a planetary gap, produced with data from \texttt{DustPy}. The top row contains simulations where the varied parameter affects not only the evolution of the dust disc but also its initial mass. Conversely, the three leftmost panels in the bottom row contains simulations where the varied parameter only influences the disc’s evolution after initialization. }
    \label{fig: Mdust vs time}
\end{figure*}

We integrate the dust surface density to obtain the total dust mass as a function of time for all three codes. A comparison of the results from different codes is shown in \autoref{fig: Mdust comparison} for a subset of simulations from the parameter study. The time evolution of the total dust mass from \texttt{DustPy} is shown in \autoref{fig: Mdust vs time} for all simulations in the parameter study. The corresponding plot with the total dust mass normalized to its initial value at $t=0$ is shown in \autoref{fig: normalized total dust mass}. 

In general, the agreement between \tripod{} and \texttt{DustPy} is good, whereas \texttt{two-pop-py} tends to produce discs that deplete their dust mass more rapidly. For \tripod{}, the largest discrepancies arise when the size distribution is poorly approximated by a power law (see \autoref{fig: size dist param}), which occurs when the densities are low (simulations $0.05-0.1M_{\ast}$, $0.02\dot{M}_0$, and $0.2r_{\rm c}$) or the fragmentation velocity is high (simulations $10-25v_{\rm frag}$). An exception is seen in simulation $0.01Z$, where all three codes agree well despite the size distribution being far from a power law. For \texttt{two-pop-py}, good agreement is achieved when turbulent diffusion is strong (simulations $5-100\alpha_{\rm t}$) or when the fragmentation velocity is significantly lower or higher than the default value of $1\, \textrm{ms}^{-1}$ (simulation $0.1v_{\rm frag}$, $10v_{\rm frag}$ and $25v_{\rm frag}$). The good agreement at high $v_{\rm frag}$ occurs despite the size distribution deviating strongly from a power law, because most of the dust mass resides near the maximum particle size, which is well reproduced by \twopop{} (see Figure~\ref{fig: particle distribution 2 sim}). Since \twopop{} evolves the dust population by tracking the largest particles, it performs well when these particles dominate the mass budget.

When varying parameters that affect the initial density - and thus the mass - of both the gas and dust disc ($M_{\ast}$, $\dot{M}_0$ and $r_{\rm c}$), simulations with lower initial dust masses deplete their dust reservoirs more rapidly (see \autoref{fig: normalized total dust mass}). This is not due to differences in $\St$, since $\St_{\rm frag} \not\propto \Sigma_{\rm gas}$, and varying the gas density therefore does not lead to a significant change in $\St$ (see \autoref{fig: st dist param}). The trend instead arises because the radial drift velocity of particles with identical $\St$ increases as the gas density decreases (see Eq. 4-5 of \citealt{Birnstiel_2012}). 
When varying $Z$, simulations with higher $Z$ deplete their dust reservoirs faster, as the increased dust-to-gas ratio leads to shorter growth timescales. When a dust-to-gas ratio of $10^{-4}$ is considered (simulation $0.01Z$), the disc loses only about 60\% of its dust mass over the full $3\, \textrm{Myr}$ evolution. 

Focusing on the three leftmost panels in the bottom row of \autoref{fig: Mdust vs time}, which shows simulations where the varied parameter only influences the disc’s evolution (i.e. not its initial mass), we find that higher $\alpha_{\rm t}$ leads to slower dust depletion. This is because increased turbulence results in smaller particle sizes (and $\St$), which leads to lower radial drift velocities. An increased fragmentation velocity allows particles to grow larger, which enhances radial drift and leads to faster dust depletion. The sharp transition observed for high $v_{\rm frag}$ is caused by an inward-moving wave of dust particles originating in the outer disc. 
Increasing the internal density of dust grains results in faster dust depletion. This is because the minimum mass on the mass grid (computed as $4/3\pi\rho_{\bullet}a_{\rm min}$) is larger, which in turn increases the smallest particle size and $\St$, giving growth a head-start.

We also performed simulations to check whether the dust mass evolution is sensitive to the initial dust size, the number of dust fluids per decade, and whether initially drifting particles are allowed. In the case of the nominal simulation, there is no significant difference; however, this could change in a different setup. 


\subsection{A planetary gap at $10\, \textrm{au}$}\label{subsec: planet gap}

\begin{figure*}
    \centering
    \includegraphics[width=\textwidth]{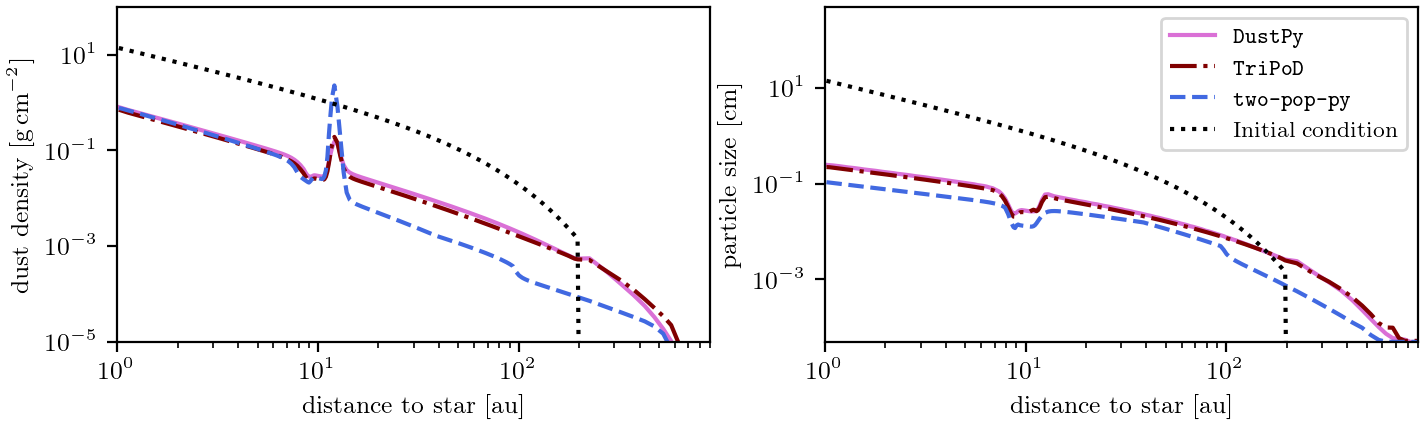}
    \includegraphics[width=\textwidth]{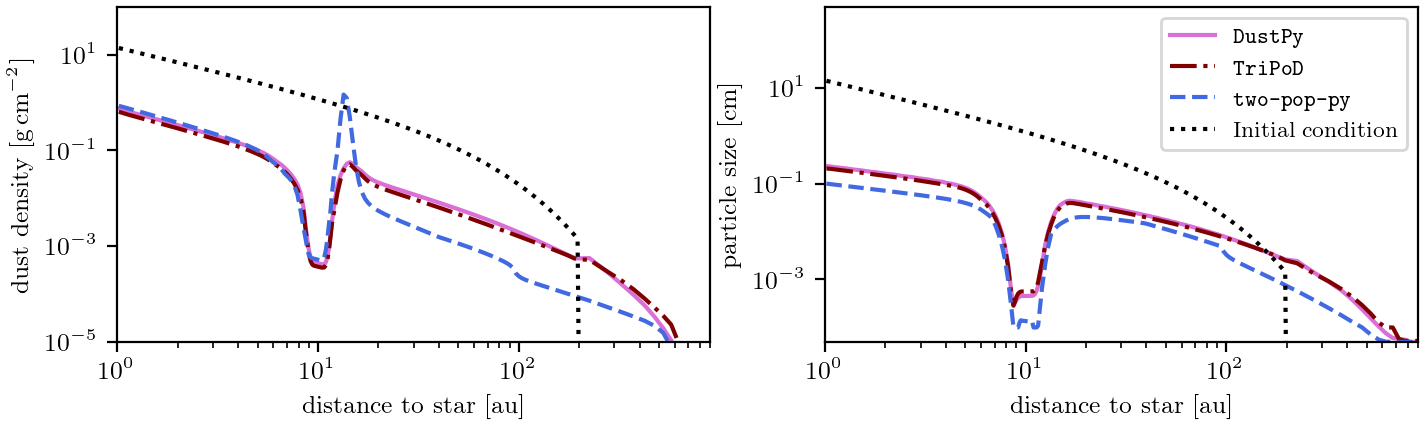}
    \caption{Comparison of the dust surface density (left) and maximum particle size (right) as a function of semimajor axis for the nominal simulation with a planet of $0.3\, \textrm{M}_{\rm jup}$ (top) and $2\, \textrm{M}_{\rm jup}$ (bottom) located at $10\, \rm{au}$. The snapshot is taken at the end of the simulation.}
    \label{fig: planetary gaps}
\end{figure*}

Two common approaches to modelling planetary gaps without explicitly including a planetary gravitational potential are (i) applying a torque on the disc or (ii) modifying the disc viscosity ($\alpha$). Here, we adopt the latter approach. We consider planetary gap profiles from \citet{Duffell2020}, and impose the inverse of the gap profile on the viscosity. 

In the nominal planetary gap set-up, we consider a Jupiter-mass planet located at $10\, \rm{au}$, and all other parameters as in the nominal simulation (see Table \ref{table: nominal}). We perform a smaller parameter study in which we vary the planetary mass between $0.1$–$2\, \rm{M}{\rm jup}$, the turbulent parameter $\alpha_{\rm t}$ between $10^{-5}$–$10^{-3}$, and the fragmentation velocity between $0.1$–$5\,\rm{m\,s}^{-1}$. We note that higher fragmentation velocities yield nonphysical results. 

\autoref{fig: planetary gaps} shows the resulting dust surface density and maximum particle size as a function of semimajor axis for two representative simulations. The most striking result is that the particle concentration at the gap edge is much stronger in \texttt{two-pop-py} than in \tripod{} or \texttt{DustPy}. This is the case in most simulations including a planetary gap. A similar result was found by Houge et al. (in prep.), who observed significantly less dust transport across planetary gaps in \texttt{two-pop-py} compared to \texttt{DustPy} when $\alpha_{\rm turb} \lesssim 10^{-3}$.
By contrast, \tripod{} and \texttt{DustPy} yield closely consistent results across the simulations.

A comparison of the total dust mass in simulations including a planetary gap is shown in \autoref{fig: Mdust comparison}. \tripod{} and \texttt{DustPy} agree closely except in the simulation with $\alpha_{\rm t}=10^{-5}$, where \tripod{} yields too high dust concentrations at the gap edge. Simulations with low fragmentation velocities or high turbulence show little to no dust concentration at the gap edge regardless of code, and the code comparison yields similar results to the corresponding no-planet simulations. The large dust masses observed for \texttt{two-pop-py} in some simulations are due to the nonphysical dust concentration at the gap edge.

The evolution of the total dust mass as a function of time from the \dustpy{} simulations is shown in the bottom right panel of \autoref{fig: Mdust vs time} for simulations with varying planetary masses. The presence of planets, regardless of their mass, does not affect the overall evolution of the total dust mass. This indicates that dust is efficiently transported across the gap, at least for the disc parameters considered in the nominal setup.  

In summary, the nonphysically high concentration of dust at the gap edge makes \texttt{two-pop-py} unsuitable for modelling dust evolution in the presence of planetary gaps, except perhaps under conditions where little or no dust pile-up is expected. \tripod{} handles planetary gaps well, except in a few edge cases.


\subsection{Interpretation of disc observations}\label{sec: disk observations}
As dust evolution codes are not only a vital part of planet formation simulations but also help us to interpret observations of discs with e.g. ALMA, in this section, we want to highlight the effect the choice of dust model has on the observables calculated from the simulations.
To model how the simulated discs would look in observations, we calculate their fluxes using the \cite{Miyake1993} scattering solution as described in \cite{Birnstiel2018}, also using the DSHARP opacities from the same work. Since the full-size distribution information is required to calculate dust emission, we have to reconstruct the size distributions for the \tripod and \twopop simulations. For \tripod this is done by reconstructing the size distribution from its parametric truncated power-law description, however \twopop assumes no particle size distribution for its algorithm, so we impose an ad hoc power-law size distribution $n(a) = a^{-q}$ truncated at $a_{\rm max}$ with $q=2.5$ if the maximal size is limited by the drift limit, $q=3.5$ in the fragmentation limit and $q=3.75$ in the drift fragmentation limit \citep{Birnstiel_2012}. Note that we omitted the simulation with the smaller minimum particle size ($a_0=10^{-5}\, \textrm{cm}$) as the opacity calculations fail for very small dust grains ($a \sim10^{-6}\, \textrm{cm}$). 

\begin{figure*}
    \centering
    \includegraphics[width=\textwidth, trim=0 0.5cm 0 0.4cm, clip]{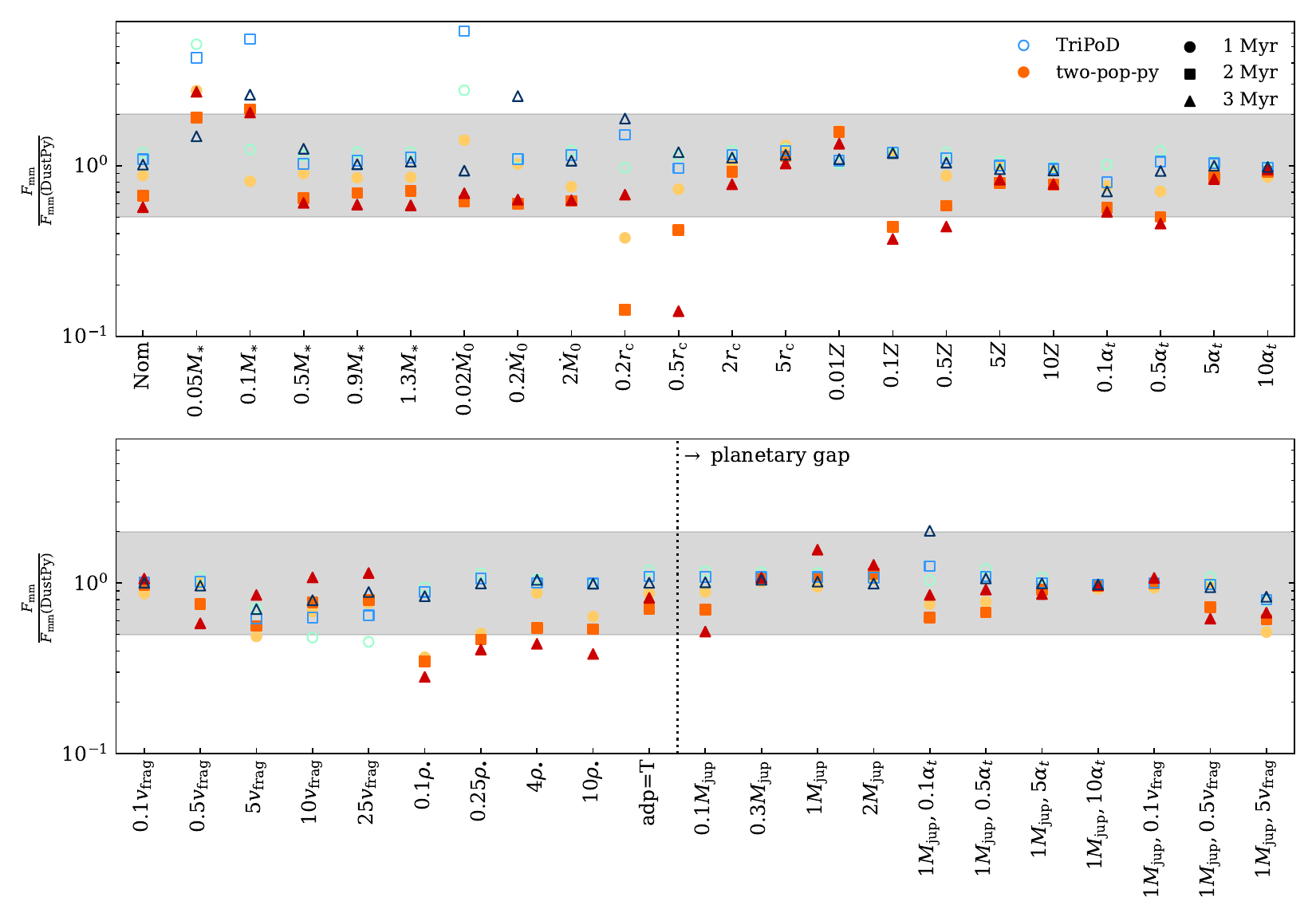}
    \caption{Same as \autoref{fig: Mdust comparison} but comparing the mm fluxes instead.}
    \label{fig: Flux vs time}
\end{figure*}

\begin{figure*}
    \centering
    \includegraphics[width=\textwidth, trim=0 0.5cm 0 0.4cm, clip]{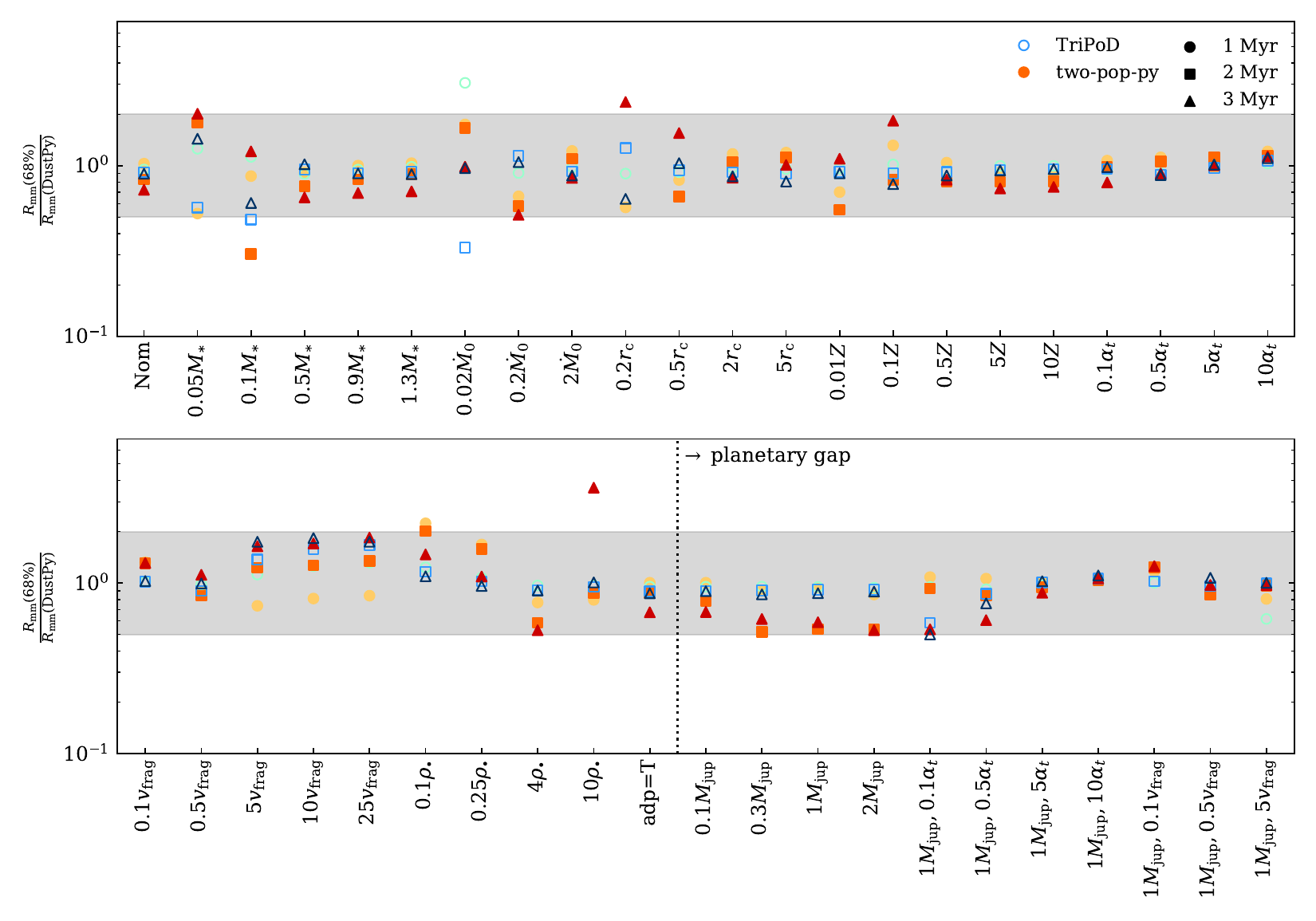}
    \caption{Same as \autoref{fig: Mdust comparison} but comparing the dust radii ($R_{\rm mm}(68\%)$) instead.}
    \label{fig: Rmm vs time}
\end{figure*}

To illustrate the result of the different models, we compare the total millimetre flux of our different setups and the outer dust radius, which is defined as the radius that encloses $68\%$ of the total flux (i.e. $F_{\rm mm}(R_{68\%}) = 0.68 \times F_{\rm tot}$). For this study, we restricted ourselves to calculating the fluxes at $1\, \textrm{mm}$ roughly corresponding to ALMA band 7. In \autoref{fig: Flux vs time}, we show the total mm fluxes of the different discs normalized to the \dustpy simulation at that time, and in \autoref{fig: Rmm vs time}, we compare the dust radii of the same simulations.

We can see the flux results unsurprisingly follow the same trends as the dust masses in \autoref{fig: Mdust comparison}; however, the \twopop models seem to be more in line with the \dustpy result in flux compared to total dust mass, indicating that our ad hoc size distribution seems to overestimate the mm flux relative to its dust mass, indicating a surplus of millimetre grains that contribute significantly to the emission. Both \tripod and \twopop show large deviations for compact discs and discs with very low mass, which is expected from size distributions and dust masses.

\begin{figure*}
    \centering
    \includegraphics[width=\textwidth]{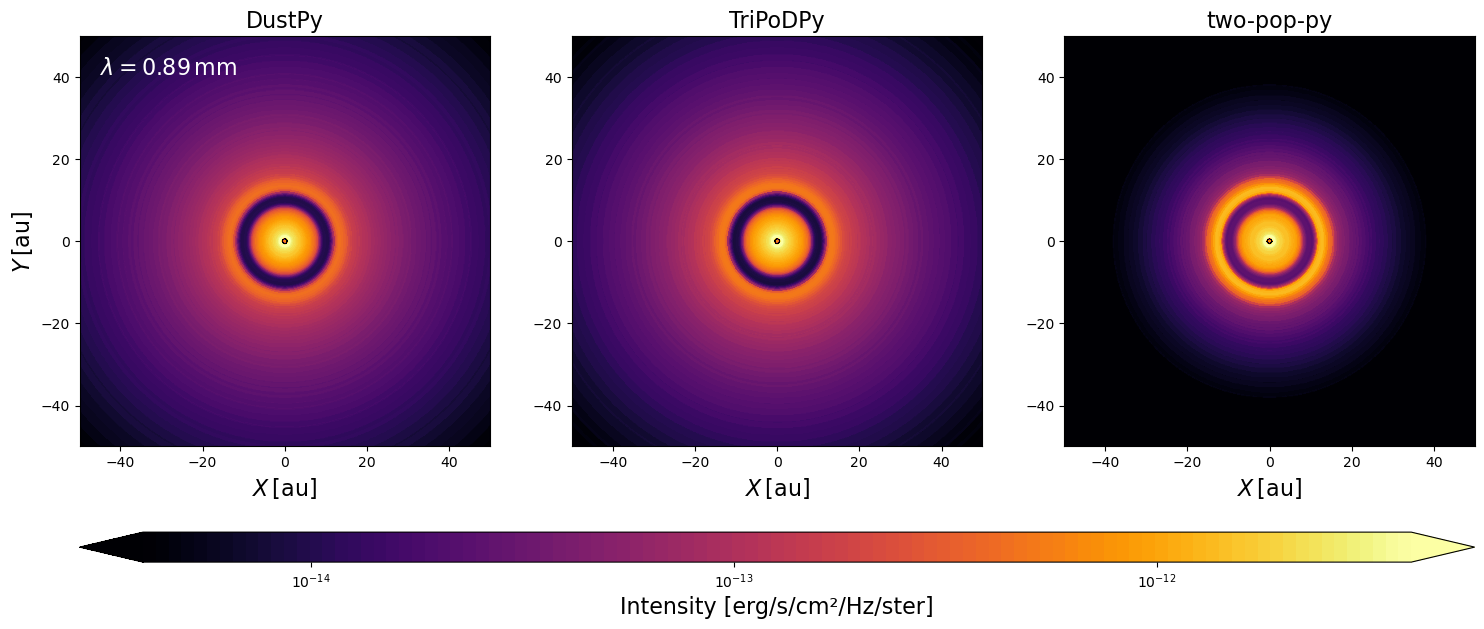}
    \caption{The intensity maps of the three different dust models of the $1\textrm{M}_{\text{jup}}$ setup at 2 Myr, created using \texttt{RADMC-3D}.}
    \label{fig:Rad_mc}
\end{figure*}

What is surprising is that the two parametric dust evolution models perform surprisingly well when including planetary gaps, which is in contradiction to the dust distributions that are very different in those cases (see \autoref{fig: planetary gaps}), as shown in Section \ref{subsec: planet gap}. To investigate this further, we run \texttt{RADMC-3D} \citep{dullemond2012} models taking the simulation snapshots from the $1 \textrm{M}_{\rm jup}$ case at 2 Myr for the three different codes, to see the radially resolved intensity (at $\lambda=0.89$mm using the default opacities). The resulting intensity maps can be seen in \autoref{fig:Rad_mc}. The Figure clearly shows that the \dustpy and \tripod{} models look very similar, whereas the \twopop emission is distributed very differently across the disc, and the fact that the total fluxes look so similar is caused by the over-bright ring at the pressure maximum compensating for the fainter outer disc.

For the dust radii, we can see that overall, the \tripod simulations seem to match the expected \dustpy results quite well, except for very low mass stars ($0.05M_\star$ and $0.1M_\star$) and low mass discs ($0.02\dot{M_0}$ and $0.2\dot{M_0}$), where we see quite a large deviation. Furthermore, some of the \twopop simulations (namely: $[5,10,25]v_{\rm frag}$ and $[0.2,0.5]r_\mathrm{c}$) show large time variation in the relative radius error as \twopop computes the growth timescale of the maximal particle size, assuming it is independent of particle size and therefore has inaccurate growth timescales in the outer disc where other sources of relative velocity make the growth timescale dependent on particle size (e.g. relative radial motion). This leads to inconsistent timing of when particles grow to the drift limit, which largely shapes the outer disc radius. For the simulations with gaps, we can see that the outer radii of \tripod{} and \twopop behave the same as in the non-gapped counterpart, with the exception being the high $v_{\text{frag}}$ runs where the gap becomes relevant to the outer dust radius.


\subsection{Planetesimal formation}

In this section, we examine how the regions of planetesimal formation are affected by various disc parameters and the choice of code. We also compare different SI criteria from the literature, as well as different approaches to defining a representative Stokes number, since most if not all SI criteria are derived for a monodisperse size distribution. We consider two different representative $\St$ for \tripod{} and \texttt{DustPy}, the density-weighted average ($\St_{\rm aver}$) and the $\St$ corresponding to the peak of the size distribution ($\St_{\rm peak}$, for \tripod{} this is the same as the maximum $\St$). For \texttt{two-pop-py} we only consider the maximum size, but we note that a density-weighted average could be calculated if the same steps are followed as in Section \ref{sec: disk observations} to reconstruct the size distribution. We note that the SI criteria L24 and L25 are derived for a global radial pressure gradient of $0.05$, whereas our simulated discs can have pressure gradients that vary with time and semimajor axis. In particular, simulations with planetary gaps exhibit strong local pressure gradients and pressure maxima. Here, we choose to ignore the dependence of the SI criteria on the pressure gradient, but emphasize that in more advanced studies, this dependence should be taken into account.

\begin{figure}
    \centering
    \includegraphics[width=\columnwidth, trim=0 0.2cm 0 0.2cm, clip]{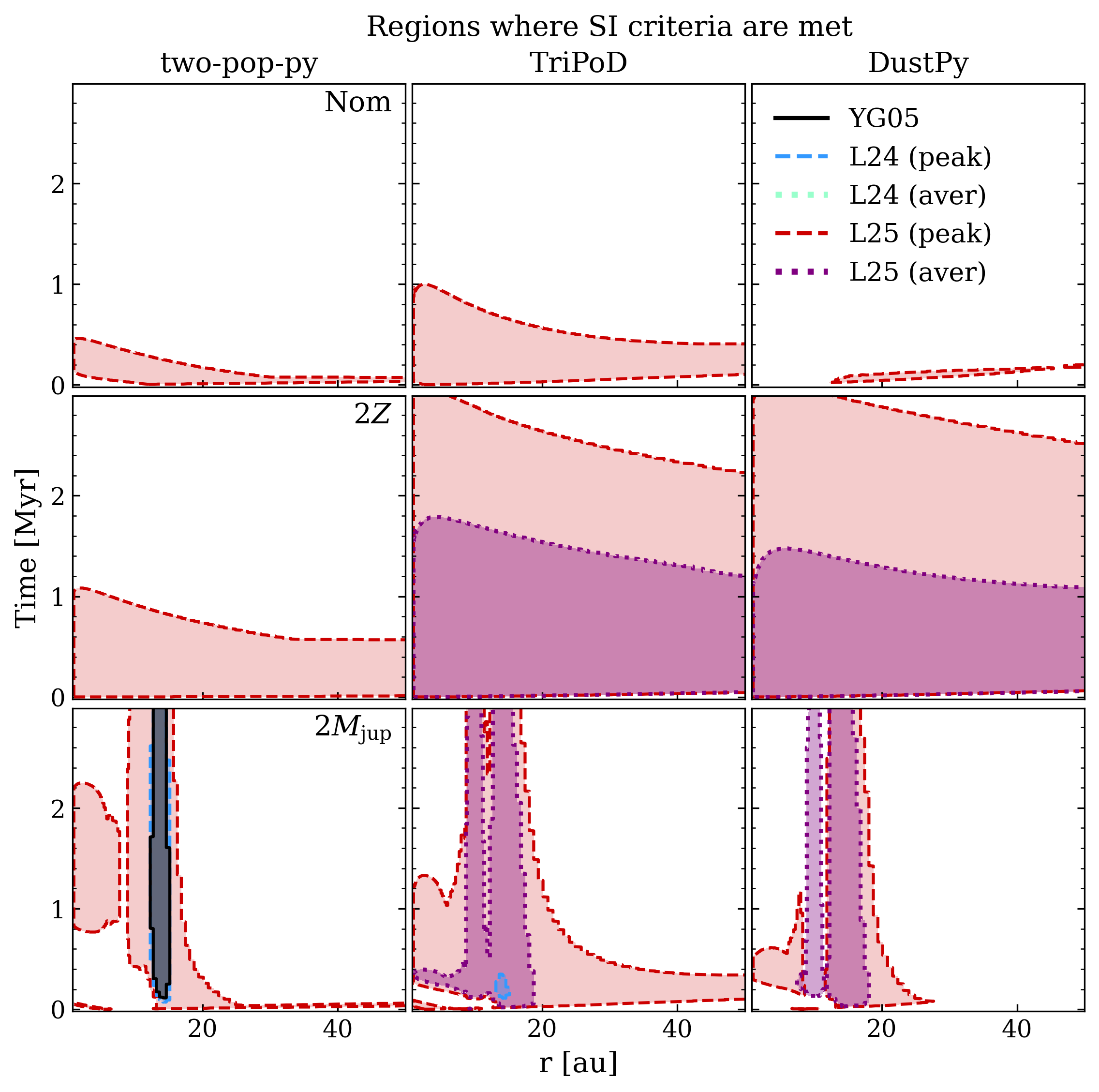}
    \caption{Comparison of regions satisfying the SI criteria for three simulations, evaluated with three 1D dust evolution codes, three SI criteria, and 2 different choices of representative $\St$ for \tripod{} and \texttt{DustPy}.}
    \label{fig: planetesimal 3 sim}
\end{figure}

\autoref{fig: planetesimal 3 sim} shows the regions on the space-time grid where the different SI criteria are met, for the three 1D dust evolution codes and three simulations. The criterion YG05, which corresponds to a midplane dust-to-gas ratio above unity, is not met in any simulation except the one with a 2-Jupiter-mass planet, and then only for \texttt{two-pop-py}, due to its nonphysically large dust concentration at the gap edge. This criterion is generally not satisfied in a smooth disc and requires significant trapping. The same holds for the criterion L24, as the required dust-to-gas ratio increases significantly when forced turbulence is taken into account. In contrast, the criterion L25 is met in various regions of the space-time grid, depending on the dust evolution code and choice of representative $\St$. 

The conditions for triggering the SI are preferentially reached near the gap edge or early during disc evolution, when the dust-to-gas ratio is highest. Planetesimal formation at planetary gap edges has been reported in previous studies (e.g. \citealt{Eriksson2020,Shibaike2020}). The exact regions where the SI criterion is met vary significantly between the codes, particularly for the nominal simulation, where the SI criterion is only marginally satisfied. \texttt{two-pop-py} predicts an end to planetesimal formation much sooner than \tripod{} and \texttt{DustPy}, which can be attributed to a more rapid depletion of dust mass. The agreement between \tripod{} and \texttt{DustPy} improves when the initial dust-to-gas ratio $Z$ is increased, allowing the SI criterion to be met with a larger margin. Representing the particle distribution by the $\St$ at its peak yields significantly larger regions where the SI criterion is met, indicating that the density-weighted average $\St$ is substantially smaller.

\begin{figure*}
    \centering
    \includegraphics[width=\textwidth, trim=0 0.2cm 0 0.2cm, clip]{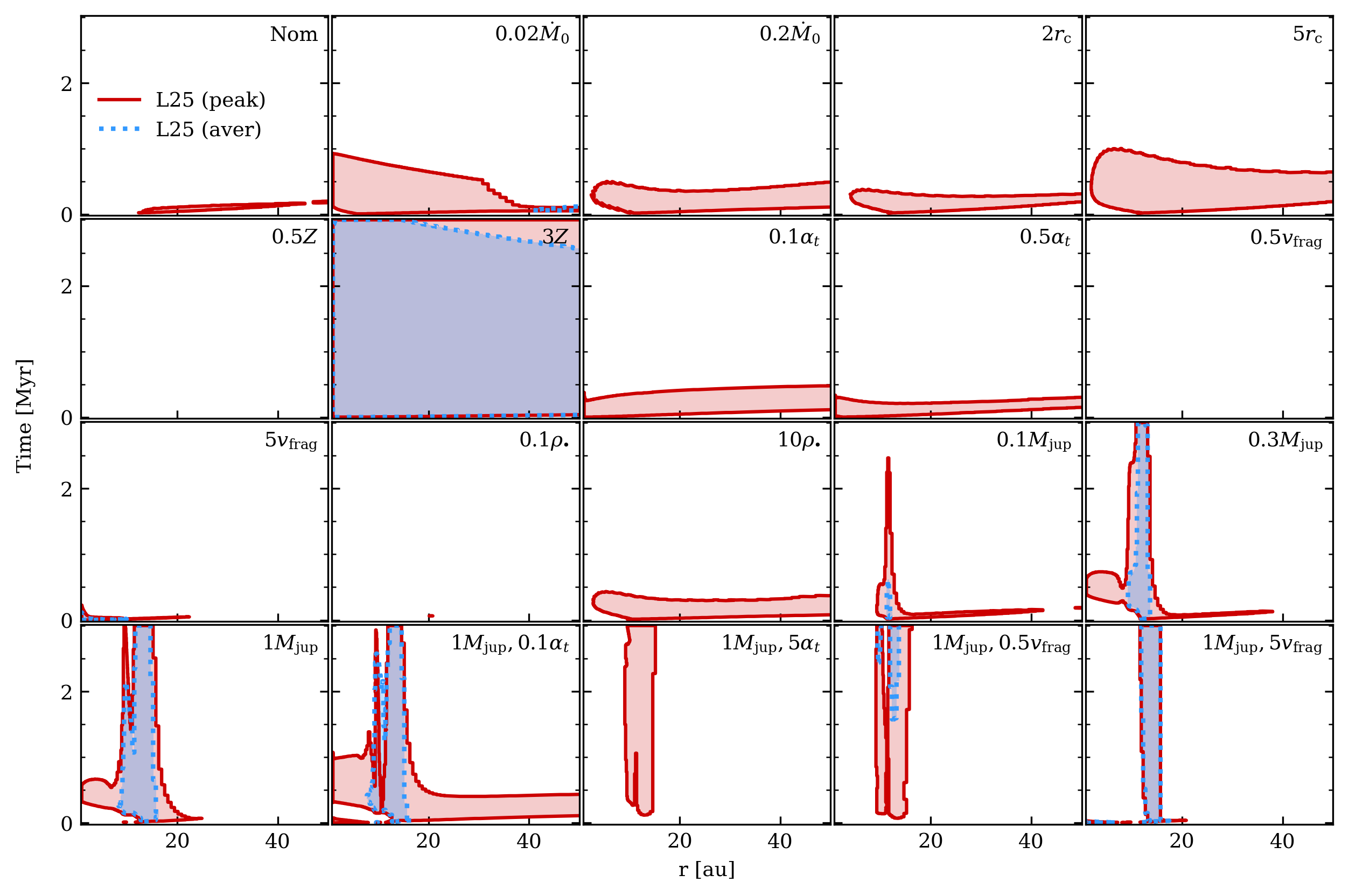}
    \caption{Regions on the space-time grid where the SI criterion L25 is met, for a subset of simulations from the parameter study. The data are produced with \texttt{DustPy}, and results are shown for both the density-weighted average $\St$ and the $\St$ at the peak of the distribution.}
    \label{fig: planetesimal param}
\end{figure*}

The regions that meet the SI criterion L25 are shown in \autoref{fig: planetesimal param} for a subset of \texttt{DustPy} simulations from the parameter study. If the density-weighted average $\St$ is used to represent the particle distribution, significant dust trapping or a supersolar dust-to-gas ratio is required for planetesimal formation to occur, regardless of other parameter choices. When considering the $\St$ at the peak of the distribution, planetesimals can form in multiple simulations, and emerging trends become apparent. Decreasing the initial dust and gas density while keeping the dust-to-gas ratio constant (decreasing $\dot{M}_0$) results in more planetesimal formation. This is mainly due to simulations with lower gas density having higher radial drift velocities, which results in an enhancement of the dust-to-gas ratio of the inner disk during the first $\sim 1\, \textrm{Myr}$, after which the outer dust disk gets depleted and the dust-to-gas ratio in the inner disk rapidly decreases. 

Increasing the exponential cutoff radius $r_{\rm c}$ leads to larger planetesimal formation regions. When $r_{\rm c}$ is small, little to no dust in the outer disc is available to replenish the inner disc, causing the dust-to-gas ratio in the inner disc to decrease rapidly over time. In contrast, when the disc is large, the inner disc continues to be replenished by dust from the outer disc, resulting in higher dust-to-gas ratios and larger regions of planetesimal formation. Increasing the turbulence level results in less planetesimal formation, as expected, since particles grow to smaller sizes and lower $\St$. No clear trend emerges for varying $v_{\rm frag}$, while increasing $\rho_{\bullet}$ results in wider planetesimal formation regions. 

The inclusion of a planet in the simulations lead to planetesimal formation near the gap edge, also in the case of relatively low-mass planets. Increasing the planetary mass results in more dust trapping at the gap edge and thus wider planetesimal formation regions. Increasing the turbulence and fragmentation velocity leads to less dust trapping and, as in the no-planet case, smaller particles which results in less planetesimal formation. When the turbulence is increased beyond $5\times10^{-4}$, no planetesimal formation is observed. We note that if the dependence of the SI criterion on the local pressure gradient was taken into account, it should be easier to form planetesimals at the gap edge, as the critical dust-to-gas ratio for clumping to occur increases with the pressure gradient \citep{BaiStone2010b}.


\subsection{Pebble accretion}

In this section, we investigate how pebble accretion-driven growth is influenced by different disc parameters and by the choice of dust evolution code. We further compare growth outcomes when assuming mono-disperse versus poly-disperse pebble accretion.
We initialize an embryo of $0.1\, \textrm{M}_{\oplus}$ at $5\, \textrm{au}$ and $0.1\, \textrm{Myr}$. We update the planetary mass using a simple Euler method with a linear time-step of $500\, \textrm{yr}$. We perform simulations where growth is stopped at the pebble isolation mass, as well as simulations where growth is allowed to continue until the end of the disc lifetime. We use equation 10-11 from \citet{Bitsch2018} for the calculation of the pebble isolation mass.

\begin{figure}
    \centering
    \includegraphics[width=\columnwidth, trim=0 0.2cm 0 0.2cm, clip]{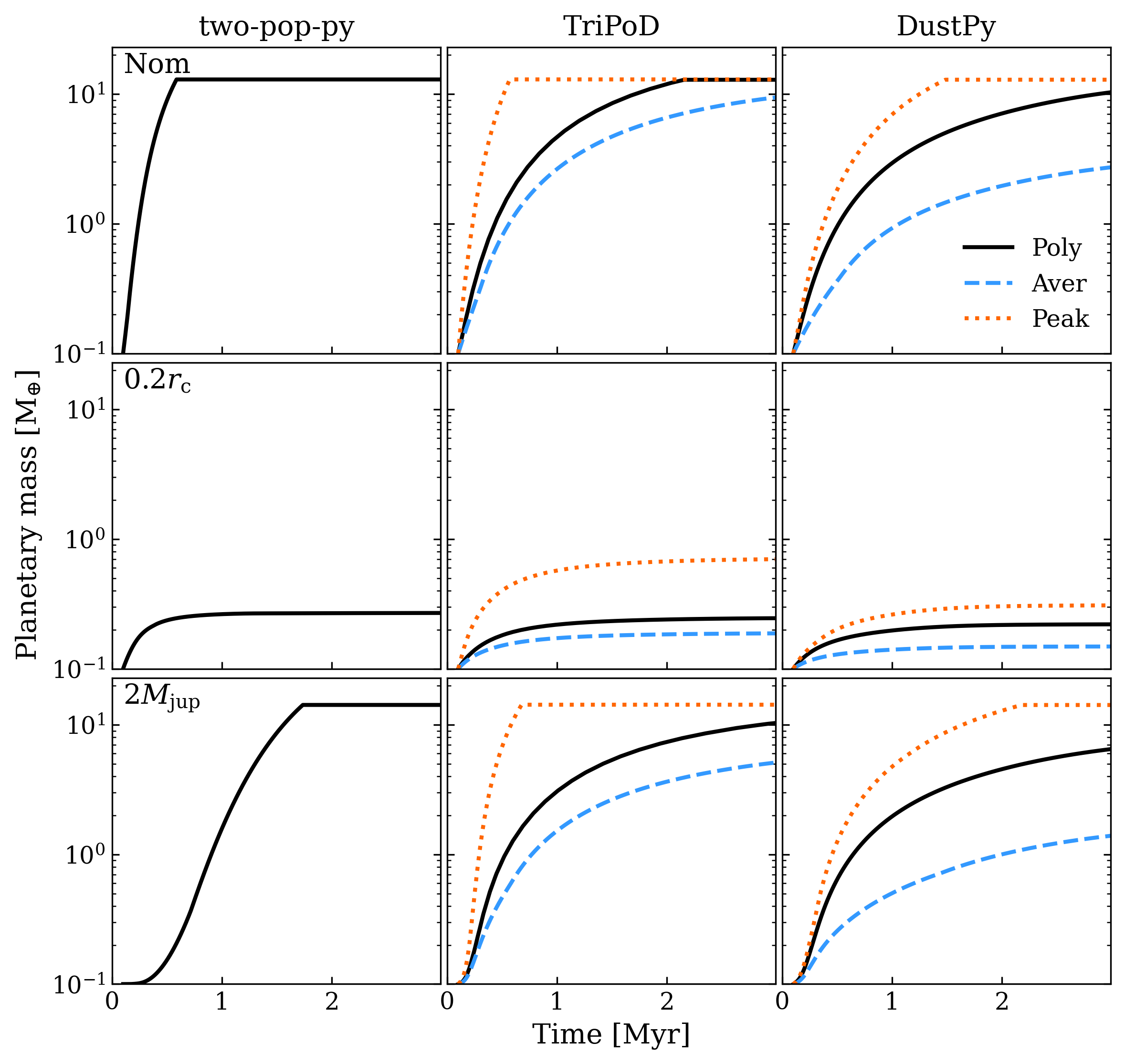}
    \caption{Pebble accretion growth tracks for planets formed at $5\, \textrm{au}$, $0.1\, \textrm{Myr}$ into disc evolution, for three different simulations and three different codes. For \tripod{} and \texttt{DustPy}, the black solid line shows the growth track assuming polydisperse pebble accretion. The blue and orange lines show the growth tracks assuming monodisperse pebble accretion, where the particle distribution is represented by the density-weighted average $\St$ and the $\St$ at the peak of the distribution, respectively.}
    \label{fig: pebble 3 sim}
\end{figure}

\autoref{fig: pebble 3 sim} shows the growth tracks of planets formed in three different simulations, produced using three different dust evolution codes. In all cases, growth was halted at the pebble isolation mass. The resulting mass evolution differs substantially between codes. When full polydisperse pebble accretion is used, the planet typically accretes more rapidly than in the monodisperse case if the distribution is represented by the density-weighted average $\St$. When considering monodisperse pebble accretion with the $\St$ at the peak of the distribution, the planet instead grows much faster than in the polydisperse case. This explains why \texttt{two-pop-py} tends to produce planets that are too massive, since the code follows the evolution of the maximum particle size.

\begin{figure*}
    \centering
    \includegraphics[width=0.98\textwidth, trim=0 0.2cm 0 0.2cm, clip]{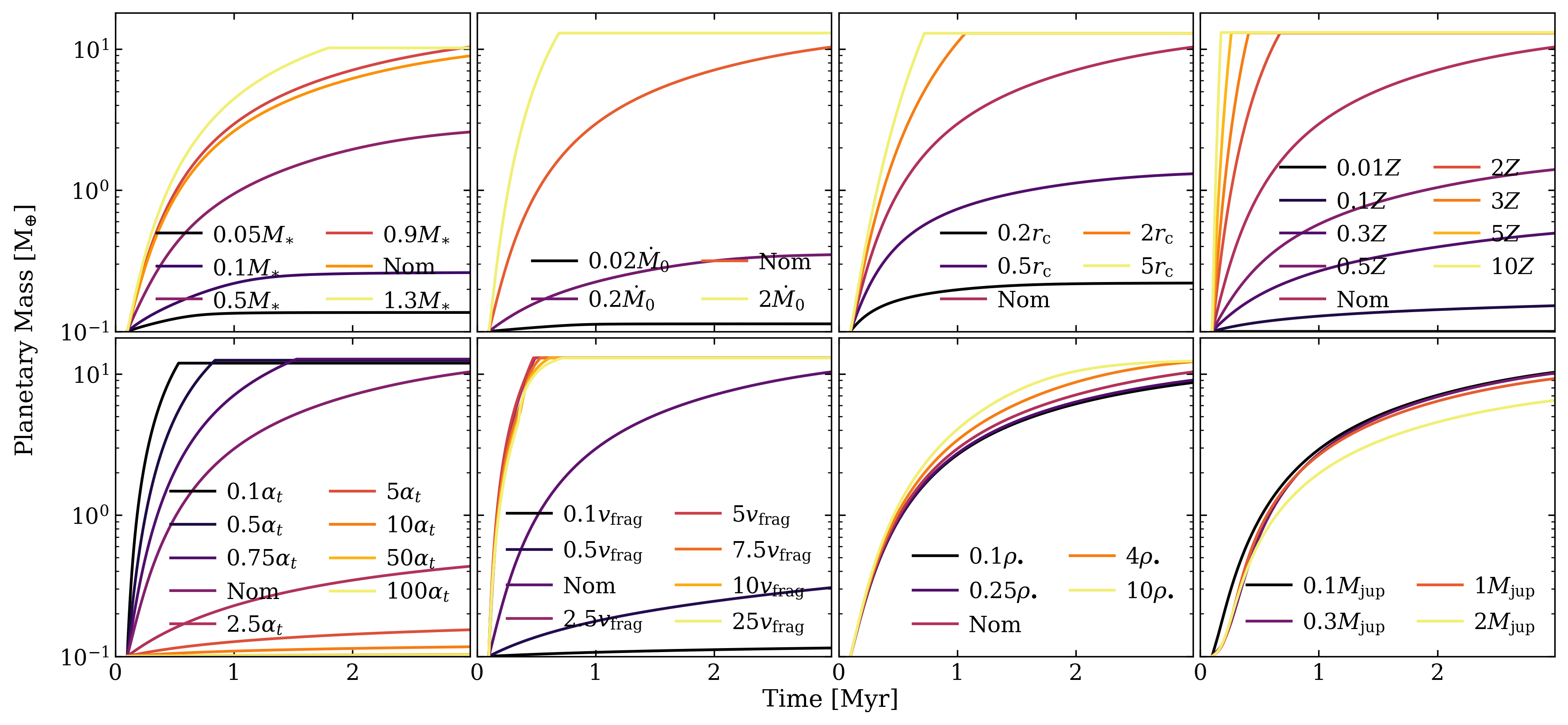}
    \caption{Pebble accretion growth tracks produced using polydisperse pebble accretion with \texttt{DustPy}, for all simulations in the parameter study. Growth is stopped at the pebble isolation mass. }
    \label{fig: pebble param DustPy}
\end{figure*}

We show the growth tracks for all simulations in the parameter study — produced using \texttt{DustPy} with polydisperse pebble accretion — in \autoref{fig: pebble param DustPy}. Increasing the stellar mass while keeping the disc-to-star mass ratio at 5\% leads to more massive planets, as higher dust densities boost the pebble accretion rate ($\dot{M}_{\rm pebb} \propto \rho_{\rm dust}$). A similar effect occurs when increasing $\dot{M}_0$ and $r_{\rm c}$. When $r_{\rm c}$ is small, growth stagnates after $\sim 1\, \textrm{Myr}$, because the pebble flux to the inner disc vanishes once dust in the outer disc is depleted. Higher dust-to-gas ratios, and thus higher dust densities, likewise produce higher accretion rates and larger planets. In contrast, stronger turbulence yields smaller particles and larger dust scale heights, leading to lower accretion rates. 

Raising the fragmentation velocity from $0.1$ to $2.5\, \textrm{m\,s}^{-1}$ accelerates growth and produces larger planets, but beyond $2.5\, \textrm{m\,s}^{-1}$ the trend reverses, as faster-drifting large particles deplete the dust disc more quickly. Increasing the internal dust density produces a modest increase in planet mass. Finally, the inclusion of a gap-opening planet at $10\, \textrm{au}$ does not significantly affect the growth of the planet at $5\, \textrm{au}$, except when the gap-opening planet has a mass two times that of Jupiter; even in that case, the final mass only decreases by $\sim 30\%$.

\begin{figure*}
    \centering
    \includegraphics[width=0.99\textwidth, trim=0 0.2cm 0 0.2cm, clip]{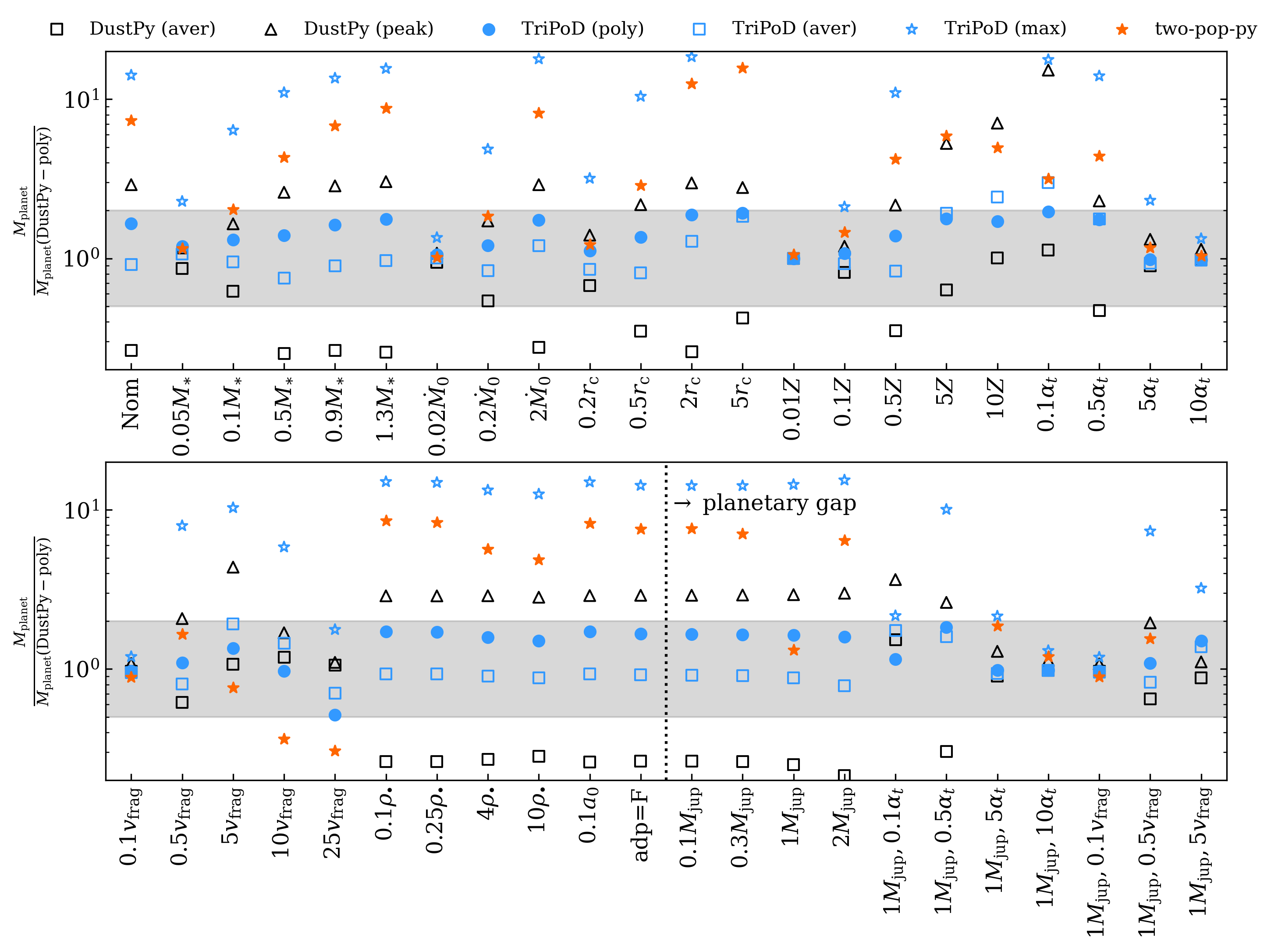}
    \caption{Comparison of the final planetary mass for a subset of simulation from the parameter study, obtained using different codes and various assumptions for the representative $\St$. Growth is not stopped at the pebble isolation mass in these simulations.  }
    \label{fig: Mtot param}
\end{figure*}

\autoref{fig: Mtot param} shows a comparison of the final planetary mass relative to the mass that is obtained when using polydisperse pebble accretion with \texttt{DustPy}, for a subset of simulations from the parameter study. For this comparison, growth is not halted at the pebble isolation mass; instead, planets continue to accrete pebbles until the end of the disc lifetime at $3\, \textrm{Myr}$. This choice is made solely for comparison purposes, as otherwise all dust codes that produce planets reaching the pebble isolation mass would yield identical outcomes, regardless of whether this occurs after e.g. $0.5, \textrm{Myr}$ or $3, \textrm{Myr}$. Because the pebble accretion rate increases with increasing planetary mass, differences between various codes and model choices increase correspondingly. For instance, the differences grow with increasing stellar mass and dust-to-gas ratio, but diminish with higher turbulence levels.

When considering polydisperse pebble accretion with \tripod{}, the final planetary mass differs by at most a factor of $\sim$ 2 compared to \texttt{DustPy}. The discrepancy is often larger than the difference in total dust mass shown in \autoref{fig: Mdust comparison}, highlighting how sensitive pebble accretion is to small variations in dust evolution. In most cases, \tripod{} produces higher planetary masses than \texttt{DustPy}. Comparing monodisperse and polydisperse accretion, both codes show significant differences. The density-weighted average $\St$ almost always yields planets of lower mass than in the polydisperse case, whereas using the peak $\St$ produces the opposite trend. \texttt{two-pop-py}, in general, almost always forms planets several times more massive than those obtained with polydisperse pebble accretion for either \tripod{} or \texttt{DustPy}.


\section{2D radial-vertical simulations}\label{sec: 2d radial-vertical simulations} The codes compared so far in this study treat protoplanetary discs in a vertically integrated way, thus ignoring vertically varying particle sizes and settling velocities.
The evolution of a dust size distribution in the direction perpendicular to the midplane (from here on referred to as the ``vertical'' direction) has been investigated in a small number of previous studies. \cite{Dullemond2005} and \cite{Tanaka2005} were among the first to report an evolution pattern known as \textit{sedimentation-driven coagulation}, in which the quickly sedimenting micron-sized dust particles in the disc's upper atmosphere coagulate on their way towards the midplane, thus accelerating their own settling. The resulting larger grains rain out, leaving only small amounts of dust in the upper atmosphere.
This process is most prevalent in systems with low levels of turbulence, where differential sedimentation becomes the dominating source of grain-grain collisions.
\cite{Zsom2011} conducted more realistic Monte Carlo simulations of this process, using a model informed by laboratory experiments.
\cite{Drazkowska2013} further investigated this process in the context of planetesimal formation.

Sedimentation driven coagulation and settling have various implications for the evolution of gas and dust in protoplanetary discs. Grain-surface chemistry is highly sensitive to the dust size distribution and the temperatures and radiation levels of the disc, both of which depend on the distance from the midplane \citep[e.g.,][]{Woidke2016}. Resulting effects like carbon depletion determine the available chemical reservoirs for planet formation and the resulting planetary atmospheres \citep[e.g.,][]{Li2021, Binkert2023}.
Stellar irradiation is mostly absorbed by small dust particles in the disc's atmosphere, whose abundance is determined by coagulation, fragmentation, and vertical mixing. Scattered stellar light in the infrared is an invaluable observational tracer of small grains and vertical mixing in protoplanetary discs \citep[e.g.,][]{Avenhaus2018,Duchene2024}.
The strength of vertical mixing can also influence the densities of larger grains in the disc midplane, which emit mostly in the sub-millimetre wavelength regime. The interpretation of such observations must therefore consider dust coagulation and opacity models \citep{Birnstiel2018}.
The amount of small grains also sets the local timescale of thermal relaxation of the gas in protoplanetary discs and therefore determines conditions for hydrodynamic instabilities \citep{Malygin2017}.

Recently, several new open-source tools have become available that make simulations of dust coagulation including the vertical direction possible. \cite{Robinson2024} introduced \texttt{cuDisc}, a GPU-based Smoluchowski solver that also includes a realistic treatment of dust dynamics. \cite{Vaikundaraman2025_mcdust} recently released \texttt{mcdust}, an open-source Monte Carlo code.
\tripod{} \citep{Pfeil2024}, although tested and calibrated in vertically integrated simulations, can also be adopted in radial-vertical models with minor modifications (see \autoref{subsec: TriPoDIntro}). 

Here, we compare some of the key features of the dust size distributions in axisymmetric, radial-vertical simulations of protoplanetary discs.
For this, we conducted two simulations per code: one at a low level of turbulence ($\alpha_\mathrm{t}=10^{-4}$), thus promoting the sedimentation-driven case, and one with a higher level of turbulence ($\alpha_\mathrm{t}=10^{-3}$), in which turbulent collision velocities and stronger vertical dust diffusion dominate the evolution of the size distribution.

To compare the dust size distributions, we interpolate (\cudisc{}) and bin (\mcdust{}) all results onto the same spherical spatial grid (taken from the \tripod{}-\texttt{Athena++} simulations).
Since \mcdust{} is a particle-based method with a finite mass resolution, plotting vertical slices is not possible, and thus radially integrated profiles must be compared. For this, we chose to divide the disc into an inner region ($<15\, \mathrm{au}$) and an outer region ($>25\, \mathrm{au}$) over which we integrate radially to achieve sufficient mass resolution for a comparison of the vertical grain size distribution.
In order to obtain a comparison of the shape of the size distributions and the grain sizes, we furthermore calculated mass-averaged particle sizes from these integrated distributions.


\subsection{Set-up}
We set up a power-law disc in all three codes with the same dust density and gas temperature at $1\,\mathrm{au}$ as in our nominal 1D simulation. All simulations have a vertically isothermal temperature structure.
The gas structure in \texttt{mcdust} is not evolving and only serves as a background for the evolving dust superparticles. In \texttt{cuDisc}, hydrostatic equilibrium is calculated numerically at every timestep based the gas column density, which we keep constant in our simulations. The temperature solver in \cudisc{} is switched off for these simulations. In order to create a static simulation with \tripod{} in \texttt{Athena++}, we initialize the simulations in hydrostatic equilibrium with zero viscosity, using an adiabatic equation of state, and without dust backreaction to avoid any perturbations that could trigger hydrodynamic instabilities.

The dust-to-gas ratio is initialized at $1\,\%$ with particles following an MRN size distribution between $5\times10^{-5} - 10^{-4}\, \mathrm{cm}$. Particles in the \mcdust{} simulations are sampled from this distribution. Since turbulent collision velocities are very low in the $\alpha_\mathrm{t}=10^{-4}$ case, we use $v_\mathrm{frag}=100\, \mathrm{cm\,s^{-1}}$ to avoid particles growing too large, thus also saving computation time in \texttt{cuDisc} with a smaller particle mass grid.
Our \tripod{} simulations are conducted in spherical geometry with a radial domain spanning from $5-50\, \mathrm{au}$ and a polar domain of $\pm 0.2 \, \mathrm{rad}$ from the midplane.
The \cudisc{} simulations are conducted on a spherical-cylindrical hybrid grid with the same polar extend and cylindrical radial boundaries. 
Since the two-dimensional simulations are considerably more expensive than our vertically integrated models, we limit the simulation time to $2\times 10^5\, \mathrm{years}$, which is sufficient to reach an equilibrium between coagulation, fragmentation, sedimentation, and vertical mixing.

We also conduct two additional \dustpy{} simulations on the same radial grid and with an identical column density structure for comparison. 
This allows us to compare the column density evolution in two-dimensional and one-dimensional simulations. In addition, \dustpy{} assumes the dust to be in constant vertical settling-mixing equilibrium, for which it derives a local dust scale height. 
Based on this parameter, the code calculates the representative settling velocities for each size bin as well as the midplane dust density, which we also use for our comparison.  

We do not investigate the dust size distributions in the drift limit in this study due to the computation-time constraints of two-dimensional simulations. We expect an outcome similar to the one-dimensional simulations, where the size distributions become more top-heavy due to the sweep-up of small grains. In this case, \tripod{}'s power-law distributions are not a very good fit to the size distribution resulting from a full coagulation simulations. However, the calibrations conducted during the development of \tripod{} and our one-dimensional comparisons have shown that a steeper power-law can reproduce the dust-mass evolution very well in this case. We note that similar dynamics occur during the initial growth phase in the upper atmosphere of the sedimentation-driven coagulation case, where large grains rain out from the upper atmosphere and sweep-up the smaller grains.


\subsection{Results}
\subsubsection{Strong turbulence}
\begin{figure*}
    \centering
    \includegraphics[width=\linewidth]{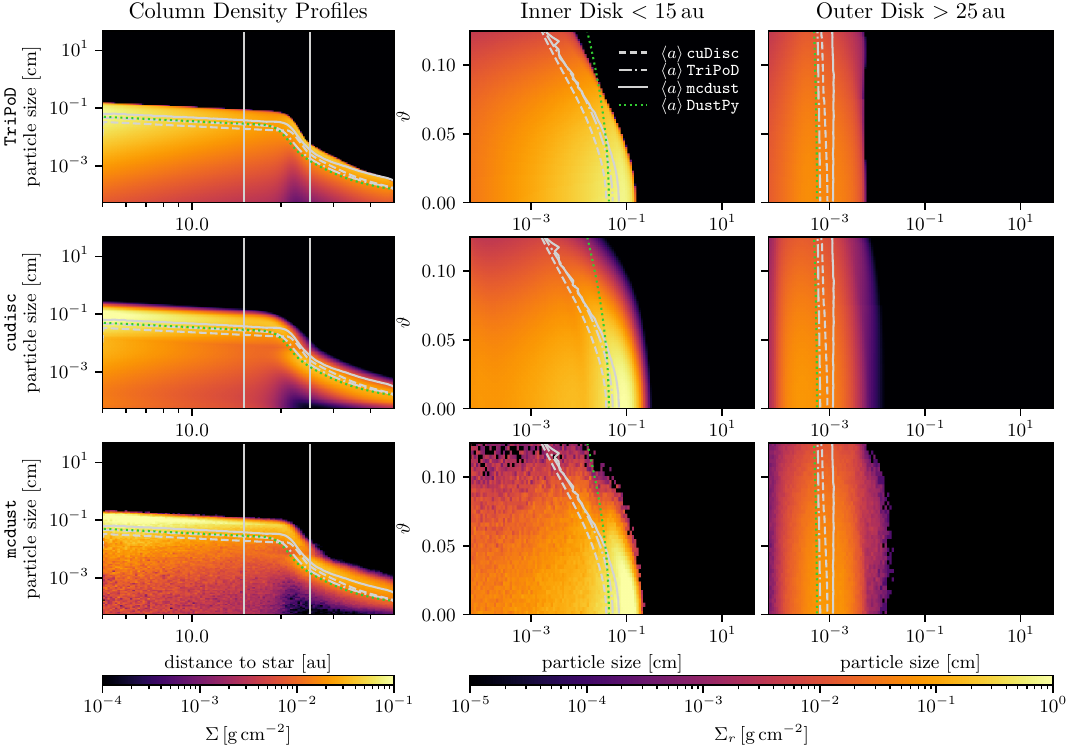}
    \caption{Comparison between the dust size distributions after $11133 \, \mathrm{yr}$ of evolution in our radial-vertical setups with a large turbulence parameter ($\alpha_\mathrm{t}=10^{-3}$) with \tripod{} (in \texttt{Athena++}), \texttt{cuDisc}, and \texttt{mcdust}. The left column shows the vertically integrated dust size distributions. The middle and right columns show the radially integrated dust size distributions as a function of polar angle in the inner disc ($r<15\, \mathrm{au}$, middle) and outer disc ($r>25\, \mathrm{au}$, right). The light grey vertical lines in the left column mark the chosen boundaries of the inner and outer disk on which the middle and right column are based. Grey lines indicate the mass-averaged particle sizes of the displayed dust size distributions. The green dotted line corresponds to a one-dimensional \dustpy{} simulation.}
    \label{fig:Full2D_alpha1e-3}
\end{figure*}
\begin{figure*}
    \centering
    \includegraphics[width=\linewidth]{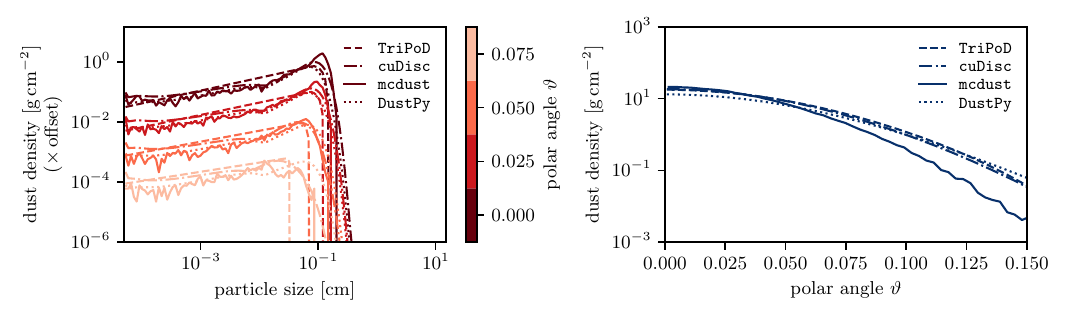}
    \caption{Radially integrated particle size distributions in the inner disc ($r<15\, \mathrm{au}$) at various distances from the disc midplane (left panel) and total vertical dust density profiles for the three codes and a large turbulence parameter of $\alpha_\mathrm{t}=10^{-3}$. Dust size distributions have been multiplied by a fixed factor for the different polar angles to improve readability.}
    \label{fig:Compare_alpha1e-3}
\end{figure*}
Turbulent particle collisions and strong vertical mixing determine the evolution of the dust size distribution in the case of large $\alpha_\mathrm{t}$ parameters. The midplane is the site of the fastest growth in this scenario.
\autoref{fig:Full2D_alpha1e-3} shows the particle size distributions of the three two-dimensional codes after $11133 \, \mathrm{yr}$. At this point, the dust has reached coagulation-fragmentation and settling-mixing equilibrium in the inner disc (i.e., $r<15\, \mathrm{au}$) but is still in the growth phase in the outer disc (i.e., $r>25\, \mathrm{au}$). 

The left column of \autoref{fig:Full2D_alpha1e-3} shows the vertically integrated dust size distributions, which overall agree well between the three codes. 
We find that the mass-averaged particle sizes of these distributions are only slightly different for the three models, with \mcdust{} showing the largest mass-averaged particle sizes and \cudisc{} the smallest. Since \cudisc{} calculates dust evolution on a size grid, more details of the size distribution can be modelled. The density distributions in \mcdust{} have a slightly enhanced peak around the fragmentation size, meaning that more mass is contained in larger particles. Hence, mass-averaged particle sizes are slightly larger in \mcdust{} compared with \tripod{} and \cudisc{}.
For comparison, we add the mass-averaged particle size from a \dustpy{} simulation with identical initial column density profile.
In this high-turbulence simulation, the mass-averaged particle sizes agree well between the one-dimensional \dustpy{} simulation and the two-dimensional simulations.

The center and right columns depict the vertical dust size distributions, radially integrated over the inner disc ($<15\, \mathrm{au}$ center column), and outer disc ($>25\, \mathrm{au}$ right column).
We compare the radially integrated vertical profiles to gain sufficient particle counts in \mcdust{} to sample the full size distribution. We split the integration at the given boundaries and take a snapshot at the given time ($11133 \, \mathrm{yr}$) to get one sample that represents the coagulation-fragmentation-settling-mixing equilibrium (inner disc) and one example of a size distribution which is still predominantly undergoing the coagulation and settling process, thus representing a more dynamical state.
Since \dustpy{} assumes continuous settling-mixing equilibrium with a vertically constant Stokes number, we can also reconstruct a vertical particle size profile for \dustpy{}, shown as green dotted lines.

All two-dimensional codes can be seen to agree well with regard to the mass-averaged particle size, shown as lines in the middle and right columns. Since Stokes numbers increase exponentially with distance from the midplane, also turbulent collision velocities increase vertically. This is why the maximum particle size decreases with altitude. 
\dustpy{} cannot account for this effect, as it assumes vertically constant Stokes numbers. The mass-averaged particle size therefore differs for the \dustpy{} simulation.
We show the radially integrated dust size distributions in the inner disc and at different distances from the midplane in the left panel of \autoref{fig:Compare_alpha1e-3}. 
At larger heights, size distributions in equilibrium (middle column) begin to slightly deviate from a simple power-law, as can be seen in the more pronounced taper in the \cudisc{} size distributions in the left panel of \autoref{fig:Compare_alpha1e-3}. These details cannot be captured by \tripod{}, which is the reason for the slightly underestimated density of large particles in the \tripod{} simulation.
\mcdust{} on the other hand begins to suffer from low particle resolution at large distances from the midplane. This becomes especially apparent when the total dust density as a function of height are compared, as in the right panel of \autoref{fig:Compare_alpha1e-3}. Here, the advantage of the fluid dynamical methods in \cudisc{} and \tripod{} becomes visible, as the accuracy of these codes does not depend on the local particle count. \mcdust{} therefore underestimates the densities at large distances from the midplane, while \cudisc{} and \tripod{} are in good agreement.
Larger particle numbers could, however, alleviate this problem in Monte-Carlo codes like \mcdust{}.
The total density profiles, shown in the right panel of \autoref{fig:Compare_alpha1e-3} agree well within $\sim\pm 2$ pressure scale heights between \tripod{} and \cudisc{}.
Since \dustpy{}'s basic assumption of settling-mixing equilibrium is fulfilled in this case, we also find very good agreement with the derived dust density profile from the \dustpy{} simulation. 
Given the large methodological differences between the three codes, the agreement in the local size distributions in equilibrium is nonetheless good; especially the slopes of the distributions agree very well for all three methods.

The more dynamical growth stage, shown in the right column of \autoref{fig:Full2D_alpha1e-3}, also shows good agreement between the three methods in the high turbulence case. Since particles are still small at this stage and since diffusion is strong, the dust has not yet sedimented. 
Since turbulent velocities determine the coagulation at almost all heights, and since densities are relatively high at large altitudes, we find a relatively constant growth pattern everywhere in the atmosphere. In \mcdust{} the growth timescale appears to be slightly underestimated. The resulting differences in the local size distributions are nonetheless small. 

Overall, the three codes capture the dynamical and the equilibrium stages of the dust evolution well in this turbulence-dominated case. Small differences can be traced back to the resolution issues of \mcdust{} and the simplifying assumptions of \tripod{} relative to \cudisc{}.


\subsubsection{Weak Turbulence}
\begin{figure*}
    \centering
    \includegraphics[width=\linewidth]{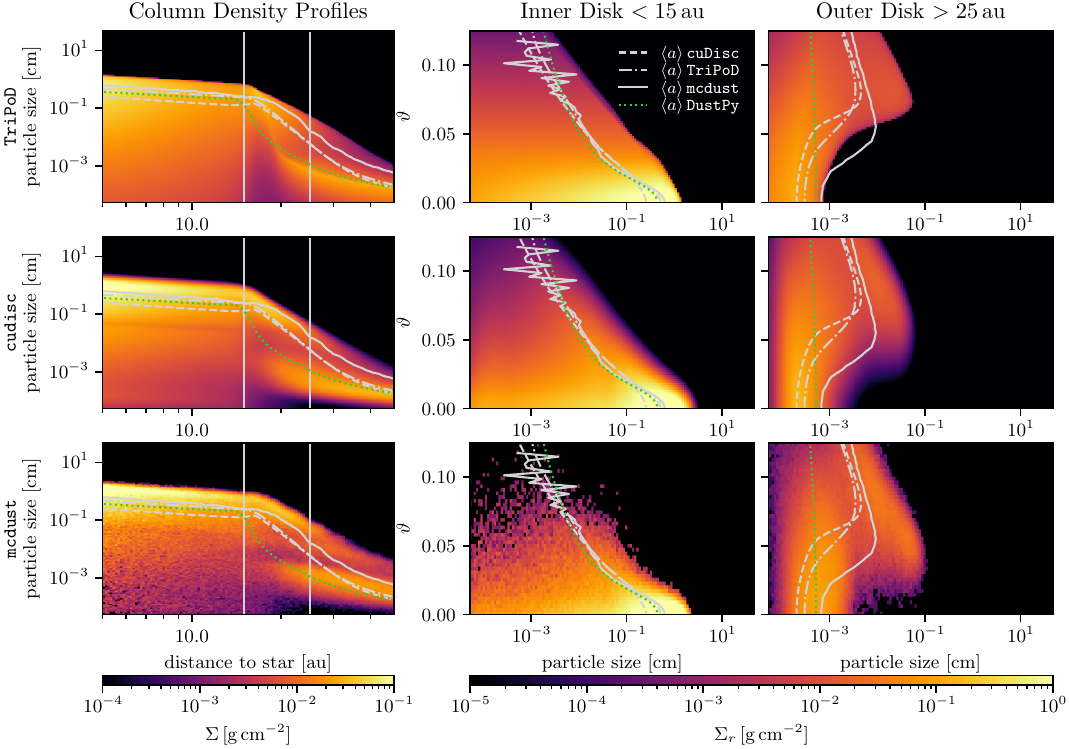}
    \caption{Comparison between the dust size distributions after $23809 \, \mathrm{yr}$ of evolution in our radial-vertical setups with a low turbulence parameter ($\alpha_\mathrm{t}=10^{-4}$) with \tripod{} (in \texttt{Athena++}), \texttt{cuDisc}, and \texttt{mcdust}. The left column shows the vertically integrated dust size distributions. The middle and right columns show the radially integrated dust size distributions as a function of polar angle in the inner disc ($r<15\, \mathrm{au}$, middle) and outer disc ($r>25\, \mathrm{au}$, right). The light grey vertical lines in the left column mark the chosen boundaries of the inner and outer disk on which the middle and right column are based. Grey lines indicate the mass-averaged particle sizes of the displayed dust size distributions. The green dotted line corresponds to a one-dimensional \dustpy{} simulation.}
    \label{fig:Full2D_alpha1e-4}
\end{figure*}
\begin{figure*}
    \centering
    \includegraphics[width=\linewidth]{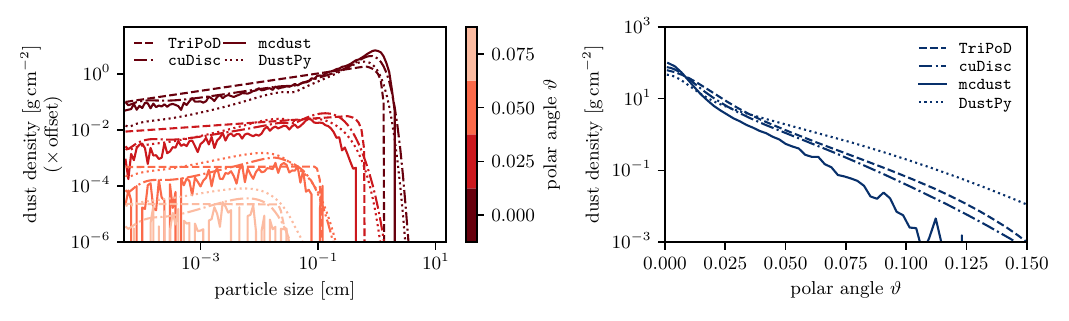}
    \caption{Radially integrated particle size distributions in the inner disc ($r<15\, \mathrm{au}$) at various distances from the disc midplane (left panel) and total vertical dust density profiles for the three codes and a low turbulence parameter of $\alpha_\mathrm{t}=10^{-4}$. Dust size distributions have been multiplied by a fixed factor for the different polar angles to improve readability.}
    \label{fig:Compare_alpha1e-4}
\end{figure*}
If turbulence is weak, particle collisions are typically dominated by relative radial and azimuthal drift and vertical settling. Sedimentation in the upper layers of the disc then oftentimes dominates over all other relative velocity components, leading to a sedimentation-driven coagulation pattern. This is the case we investigate in our simulations with a low turbulence parameter of $\alpha_\mathrm{t}=10^{-4}$, shown in \autoref{fig:Full2D_alpha1e-4} and \autoref{fig:Compare_alpha1e-4}. 
Since the particles can grow to larger sizes in these simulations due to the low level of turbulence, we choose a later snapshot at $23809 \, \mathrm{yr}$ to capture the equilibrium state of the inner disc and the dynamical evolution of the outer disc.

The left panel of \autoref{fig:Full2D_alpha1e-4} again shows the vertically integrated dust size distributions for the three codes, with lines representing the mass-averaged particle sizes.
Because of the lower level of turbulence, particles can now grow to larger sizes. Collision velocities are now dominated by drift and, in the upper layers, settling.
The growth patterns can again be seen to agree well between the three two-dimensional codes. \dustpy{}, however, which cannot account for the sedimentation-driven-coagulation in the upper layers, underestimates the mass-averaged particle size in the growth phase.  

The middle panel of \autoref{fig:Full2D_alpha1e-4} depicts the radially integrated size distributions as a function of height in the inner disc. In these regions, growth, settling, and mixing have already reached an equilibrium state. 
Due to the larger particle sizes and enhanced sedimentation, we can now see that \mcdust{} begins to suffer from low mass resolution in the upper layers.
The mass-averaged particle sizes, however, still agree very well between the three codes and also with the settling-mixing equilibrium of \dustpy{}.
In the midplane, centimetre-sized particles can be found in the inner disc regions, while the upper atmosphere is composed of sub-mm particles. 
As fragmentation is now dominated by relative settling and radial drift, we also find the size distributions in \cudisc{} and \tripod{} to follow shallower slopes. This behavior was already discussed previously \citep{Birnstiel2011, Pfeil2024, Pfeil2025}, and results in an increased abundance of smaller grains. 
These flatter size distributions increase the mass resolution problem of \mcdust{}, as small grains dominate the particle count even more compared to the turbulence-driven case, despite the increase in maximum grain size.

The right panel of \autoref{fig:Full2D_alpha1e-4} shows the dynamical evolution of the size distribution in the outer regions, where growth has not yet reached the fragmentation limit. 
The growth pattern is now different from the high turbulence case, as grain collisions are dominated by relative settling velocities in the upper disc layers. Since these collisions are much more frequent in the upper layers of the disc, due to the higher settling velocities, we can observe the typical sedimentation-driven coagulation pattern. Starting from large altitudes, grains grow and sediment---sweeping up smaller particles during their falls, which further accelerates their growth. 
A wave of large particles rains out into the disc's midplane, leaving the upper layers relatively depleted of dust. 
Most mass is still concentrated in the disc's midplane, where turbulence-dominated coagulation proceeds. This is visible as a particle overdensity in the vertically integrated dust size distributions at $\sim 10\, \mathrm{\mu m}$ between $20-50\, \mathrm{au}$ in the left column of \autoref{fig:Full2D_alpha1e-4}. The accelerated growth in the atmosphere is clearly visible as a second population at sizes larger than $\sim 10\, \mathrm{\mu m}$ in these panels. This means that the vertically integrated size distribution in the dynamical stages of sedimentation-driven coagulation is bimodal, showing the distinct dynamics of the disc's midplane and atmosphere.
This pattern is reproduced in all three codes. 
\dustpy{}, however, cannot account for this vertically varying growth pattern and underestimates the particle sizes in the growth phase. This is visible in the underestimated mass-averaged particle sizes in the left and right panels of \autoref{fig:Full2D_alpha1e-4}.

In \autoref{fig:Compare_alpha1e-4}, we show the radially integrated size distributions in the inner disc as a function of height in the left panel. As can be seen, the \mcdust{} size distributions are poorly sampled at large distances from the midplane and for small particle sizes.
Despite this shortcoming, all three codes agree reasonably well for this setup closer to the midplane. 
The \dustpy{} size distributions show similar slopes. Densities are, however, overestimated in the \dustpy{} simulation and underestimated in \mcdust{} compared to \cudisc{} and \tripod{}.


\section{Discussion}\label{sec: discussion}


\subsection{Dust evolution codes not included in this study}
The dust evolution codes that are part of this code comparison are \texttt{DustPy}, \tripod{}, \texttt{two-pop-py}, \texttt{cuDisc}, and \texttt{mcdust}. In this section, we discuss a few additional open-source dust evolution codes used in the literature. 

The \texttt{pebble-predictor} code by \citet{Drazkowska2021} is an open-source semi-analytic model for predicting the flux-averaged Stokes number and total flux of pebbles as a function of time and semimajor axis in an unperturbed disc. Unlike in \texttt{two-pop-py}, the evolution of the gas surface density is neglected in \texttt{pebble-predictor}, and the evolution of the dust surface density is approximated by keeping track of the mass budget. For this reason, this code is not suitable for use in studies in which the gas disc undergoes significant evolution.

The code \texttt{PHANTOM} by \citet{Price2018} is an open-source smoothed particle hydrodynamics (SPH) code that includes a module for particle growth and fragmentation \citep{Vericel2021}. A key approximation of this module is that they consider locally monodisperse size distributions, such that each SPH particle contains a single size that evolves due to the local disc conditions. Because all particles are assumed to have the same size locally, there is no relative velocity due to drag, and they consider only the relative velocity due to turbulence. Furthermore, coagulation is always between equal-sized grains and result in perfect merging. For fragmentation, the code provides either smooth fragmentation (mass loss increases with relative velocity) or hard fragmentation (most mass is lost regardless of relative velocity) as options. The size of the resulting fragment is obtained by equating the relative velocity with the fragmentation velocity. Phantom includes both a two-fluid and a one-fluid algorithm (relevant only for the most strongly coupled particles), and the growth module is implemented in both. Because the method assumes that the local size distribution is highly peaked around a certain value, this method is unsuitable for shallow size distributions. 

\citet{Lebreuilly2023} developed the \texttt{SHARK} code. \texttt{SHARK} follows gas and dust hydrodynamics while accounting for grain evolution by numerically solving the Smoluchowski equation. It accounts for coagulation and fragmentation with various collision sources such as turbulence, Brownian motion, hydrodynamical drift and ambipoar drift. Initially a 1D code, it was extended to 2D calculations to study the streaming instability \citep{Vallucci-Goy2026}. In these simulations, a monodisperse approach of dust growth was employed assuming a turbulent collisional velocity. The code can also be used using evolutionary tracks of tracer particles obtained from 3D simulations to solve for coagulation in a post-processing step. This was done by \cite{Navarro-Almaida2024} to follow the chemical evolution of a collapsing prestellar core.

\cite{Marchand2021, Marchand2022, Marchand2023} developed \texttt{Ishinisan}, an analytical method to track dust coagulation via a reduced variable $\chi$ that tracks the grain evolution. Although this method allows for very inexpensive calculation of the initial growth phases of dust grains, it cannot account for any other physical processes altering the size distributions, such as fragmentation.
It is thus mostly suited for studies of the initial phases of dust growth, such as grain evolution in the early stages of protostellar collapse where fragmentation is likely not yet occurring.

The \texttt{Pencil Code} \citep{BrandenburgDobler2002} is an open-source, high-order finite-difference code for hydrodynamics and MHD. The code has a dust coagulation module that is similar to the one used in the \texttt{PLANETESYS} code in \citet{Eriksson2020}. Dust is treated as superparticles, and collisions between superparticles are modelled using a Monte Carlo approach in which collisions can result in coagulation, fragmentation, erosion, or bouncing. One major limitation of this model is the treatment of fragmentation: if two particles collide and fragment, the mass of the larger particle is set equal to that of the smaller particle. As a result, small dust is not replenished, which, for example, affects the transport of dust across planetary gaps.

\subsection{Effects not included in the models}
This study has assumed relatively simple simulation setups. Some of the more important processes not taken into account include bouncing, grain composition, and icelines.
Studies of bouncing suggest that it occurs at velocities lower than the fragmentation velocity \citep{Guttler2010}. \citet{DominikDullemond2024} implemented a formalism for the bouncing barrier into \texttt{DustPy} and found that it led to smaller maximum particle sizes and more monodisperse size distributions, almost depleted of $\mu\textrm{m}$-sized grains. As both planetesimal formation and pebble accretion depend strongly on particle size, the inclusion of bouncing would thus have a significant impact on planet formation simulations. The change in the size distribution, particularly the depletion of small particles, would furthermore affect how discs appear in observations.

In our simulations, we considered dust grains to have identical material properties throughout the disc. In reality, solid particles in the disc are composed of a mixture of dust and ice, with the ice composition varying with the surrounding gas temperature. These ice-coated particles may exhibit different sticking properties depending on their composition and location in the disc, although this remains under debate \citep{GundlachBlum2015, MusiolikWurm2019}. Even if this remains uncertain, the composition still plays an important role, as ice on the particle surfaces evaporates when drifting particles cross various icelines. When ice evaporates in regions where it cannot recondense, the local solid density decreases. The cycle of evaporation, outward diffusion, and recondensation of water vapor around the snow line has furthermore been identified as a promising site for planetesimal formation via the streaming instability, as it leads to a pile-up of large particles outside the snow line \citep{DrazkowskaAlibert2017, SchoonenbergOrmel2017}. 

As briefly mentioned in Section \ref{subsec: 1D parameter choices}, collisions between dust aggregates likely result in some degree of compaction. The outcome of dust collisions depends on the porosity of the particles \citep{Guttler2010}, and taking compaction into account could therefore have a significant effect on dust evolution. Examples of dust evolution studies that include compaction can be found in \citet{GarciaGonzalez2020} and \citet{Michoulier2024}.

This study assumes a simplified model for the evolution of the underlying gas disc. In viscously driven accretion, angular momentum is transported radially outward in the disc, causing the disc to expand over time. However, due to a lack of viable mechanisms to drive this angular momentum transport, the direct removal of angular momentum vertically by magnetized disc winds provides another plausible cause of accretion. Nevertheless, all 1D radial simulations presented here adopt a viscous accretion disc model. There are two main reasons for this: (i) the default versions of the dust evolution codes all assume that the gas evolves viscously; and (ii) viscous accretion discs remain the most commonly used model in planet formation simulations. Significant changes to the gas disc evolution would certainly affect dust evolution, although it is nontrivial to predict exactly how.

Finally, the default versions of the 1D radial dust evolution codes do not include a mechanism for disc dispersal, which could occur, for example, through photoevaporation. The inclusion of photoevaporation would significantly alter the dust evolution toward the end of the disc’s lifetime when the gas density becomes low. In the case of planetesimal formation, the increased dust-to-gas ratio caused by the drop in gas density could trigger the formation of a late generation of planetesimals \citep{Carrera2017}.


\section{Conclusion}\label{sec: conclusion}

In this study, we compared several open-source dust evolution codes to determine whether the choice of code influences the outcome of disc evolution and planet formation simulations. We also conducted a parameter study to examine how key parameters affect disc evolution and planet formation.
Our main findings are as follows:
\subsection*{1D Comparison}
Our 1D radial simulations of unperturbed discs using \texttt{DustPy}, \tripod{}, and \texttt{two-pop-py} show that \texttt{DustPy} and \tripod{} generally agree well, whereas \texttt{two-pop-py} tends to deplete the disc dust mass too rapidly. \tripod{} deviates from \texttt{DustPy} when the size distribution is poorly approximated by a power law, which occurs, for example, when gas densities are low or fragmentation velocities are high.
    \paragraph*{Planet-induced Gaps:} In simulations including planetary gaps, dust concentrations at the gap edge are much higher in \texttt{two-pop-py} than in \texttt{DustPy} and \tripod{}, which closely agree except in some edge cases. This demonstrates that \texttt{two-pop-py} is unsuitable for modelling discs with planetary gaps.
    \paragraph*{Observables:} We calculated millimetre fluxes and dust radii for all simulated discs to assess how the choice of code affects the interpretation of disc observations. Across a wide range of physical parameters, the three 1D codes typically produce results that agree within 50\%, with larger deviations appearing mostly at low gas densities. However, the radially resolved intensities vary significantly in the cases with planetary gaps, where \texttt{two-pop-py} significantly overestimates the brightness of the resulting ring.
    \paragraph*{Planetesimal formation via SI:} Where in time and space the streaming instability criteria are met varies significantly with the choice of code, particularly when the criteria are just marginally met. The differences diminish when the SI criteria are strongly satisfied (e.g., at high dust-to-gas ratios), although \texttt{two-pop-py} consistently predicts an earlier end to planetesimal formation due to faster dust depletion. All three 1D codes predict planetesimal formation at planetary gap edges, provided that the amount of turbulence is not too high.
    \paragraph*{Pebble Accretion:} Pebble accretion is sensitive both to the chosen dust evolution code and to whether monodisperse or polydisperse accretion is used. Polydisperse pebble accretion with \tripod{} produces planets of similar mass to those from \texttt{DustPy}. Using \texttt{two-pop-py}’s single-particle output, however, yields planets several times more massive. Adopting monodisperse pebble accretion together with the density-weighted average particle size in \texttt{DustPy} results in significantly lower mass planets than in the polydisperse case, whereas using the peak size produces the opposite trend. The differences between mono- and polydisperse pebble accretion are smaller for \tripod{} due to its smoother size distribution. 
\subsection*{2D Comparison}
The models including the vertical direction of the disc included in our study generally agree well with each other and reproduce the general growth patterns under the influence of dust sedimentation. 
    While full coagulation solvers for the dust fluid approach like \cudisc{} provide a rigorous numerical treatment of dust coagulation, they are too expensive to be coupled with hydrodynamic solvers. Monte Carlo codes like \mcdust{} provide another framework for dust coagulation with unique advantages, like the easily available tracking of individual representative particles. These particle-based methods are, however, computationally expensive and suffer from low resolution in the upper disc atmosphere. They can thus also not be efficiently coupled to hydrodynamic solvers. Approximate methods like \tripod{} provide the simplicity necessary for the coupling of dust coagulation and hydrodynamics. When properly tested and calibrated, they can reproduce dust densities and growth patterns of full Smoluchowski solvers. \tripod{}, however, is limited to the modelling of power-law dust size distributions and therefore cannot reproduce the finer details of the size distributions or any kind of multi-modal size distribution, as might be present at large altitudes.
    What method is appropriate for a given application is therefore highly dependent on the relevant physics that are to be modelled and the regions within the disc that are of interest. 
    Hydrodynamic modelling requires fast and simple dust coagulation prescriptions like \tripod{}. Accurate radiative transfer depends on the smallest available particles, potentially at large distance from the disc midplane. If hydrostatic disc structures are to be calculated, \cudisc{} is thus the tool of choice. 
    If the tracking of individual particles are necessary to model chemistry, composition, and other dust properties, Monte Carlo method like \mcdust{} are well suited. 

\section*{Author contributions}
LE conceived and led the project, performed the 1D simulations (except those including planetary gaps), carried out the corresponding analysis, and led the writing of the manuscript. TP carried out the 2D simulations with \tripod{} and \cudisc{}, led the analysis and writing of the section on radial-vertical simulations, and performed the 1D simulations including planetary gaps. NK performed the simulated observations and led the analysis and writing of that section. VV carried out the 2D simulations with \mcdust{} and contributed to the description and interpretation of those results. All authors contributed to the model set-ups, scientific discussion, and writing of the manuscript.


\section*{Acknowledgements}
We thank S. Stammler and T. Birnstiel for useful discussions and contributions to the model setup, and P. Armitage and U. Lebreuilly for helpful discussions. We also thank the referee for their useful comments. 
LE acknowledges the support from NASA via the Emerging Worlds program (\#80NSSC25K7117). VV acknowledges funding from the European Union under the European Union’s Horizon Europe Research \& Innovation Programme 101040037 (PLANETOIDS). NK acknowledges funding from the European Union under the European Union’s Horizon Europe Research and Innovation Programme 101124282 (EARLYBIRD) Views and opinions expressed are,  however, those of the author(s) only and do not necessarily reflect those of the European Union or the European Research Council. Neither the European Union nor the granting authority can be held responsible for them.

\section*{Data Availability}
All scripts used for initialising, running, and analysing the simulations are open source and available in the following GitHub repository: \url{https://github.com/astrolinn/dustComparison.git}. Selected data snapshots from the dust evolution simulations are publicly available at \url{https://doi.org/10.5531/sd.astro.11}. The full dataset underlying this article is available from the corresponding author upon reasonable request.



\bibliographystyle{mnras}
\bibliography{refs}




\appendix

\section{Additional plots}
\label{sec: additional plots}



\begin{figure*}
    \centering
    \includegraphics[width=\textwidth]{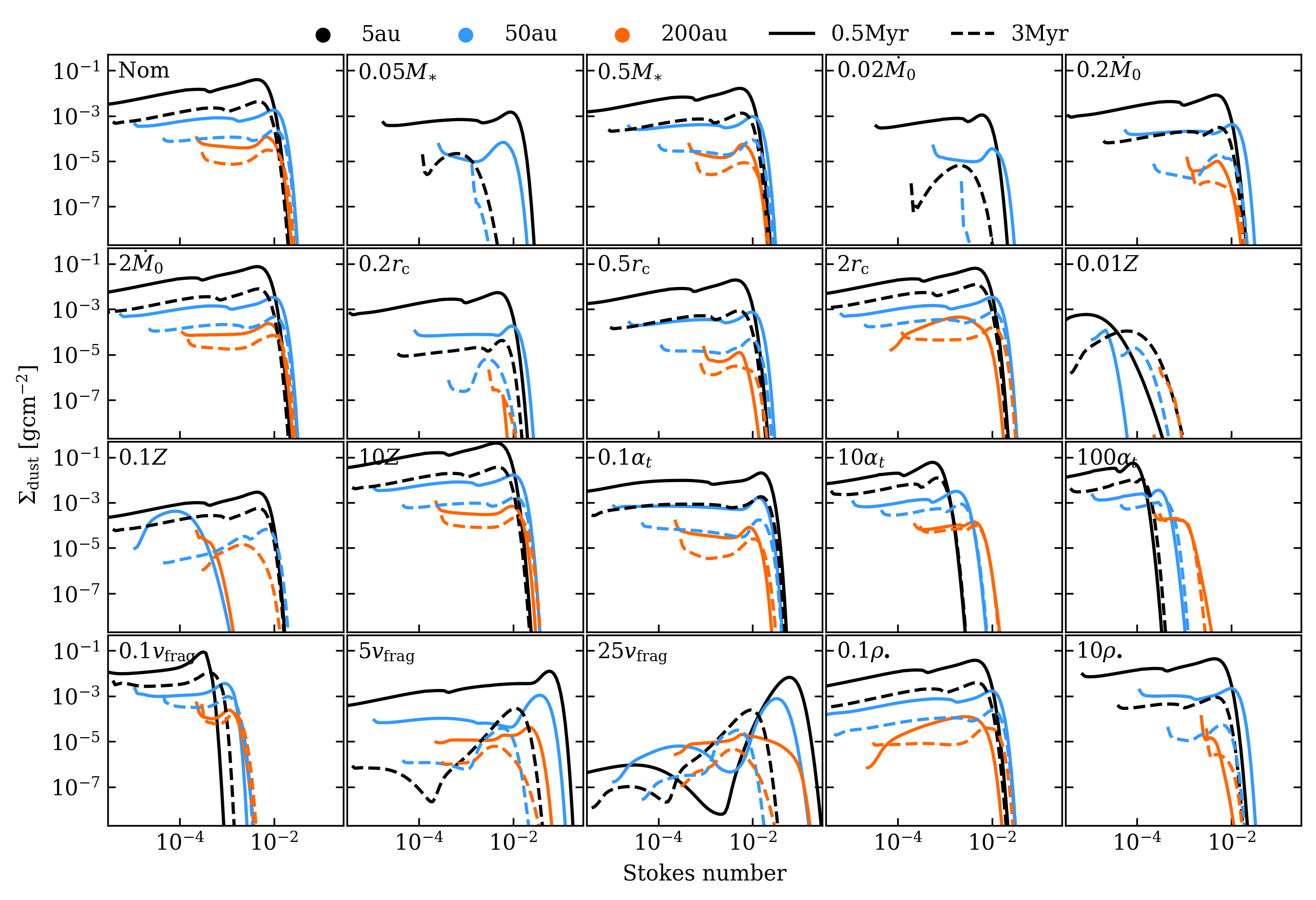}
    \caption{Stokes number distribution for a subset of simulations from the parameter study, produced using \texttt{DustPy} and shown for three different semimajor axes and two different times. }
    \label{fig: st dist param}
\end{figure*}

\begin{figure*}
    \includegraphics[width=\textwidth]{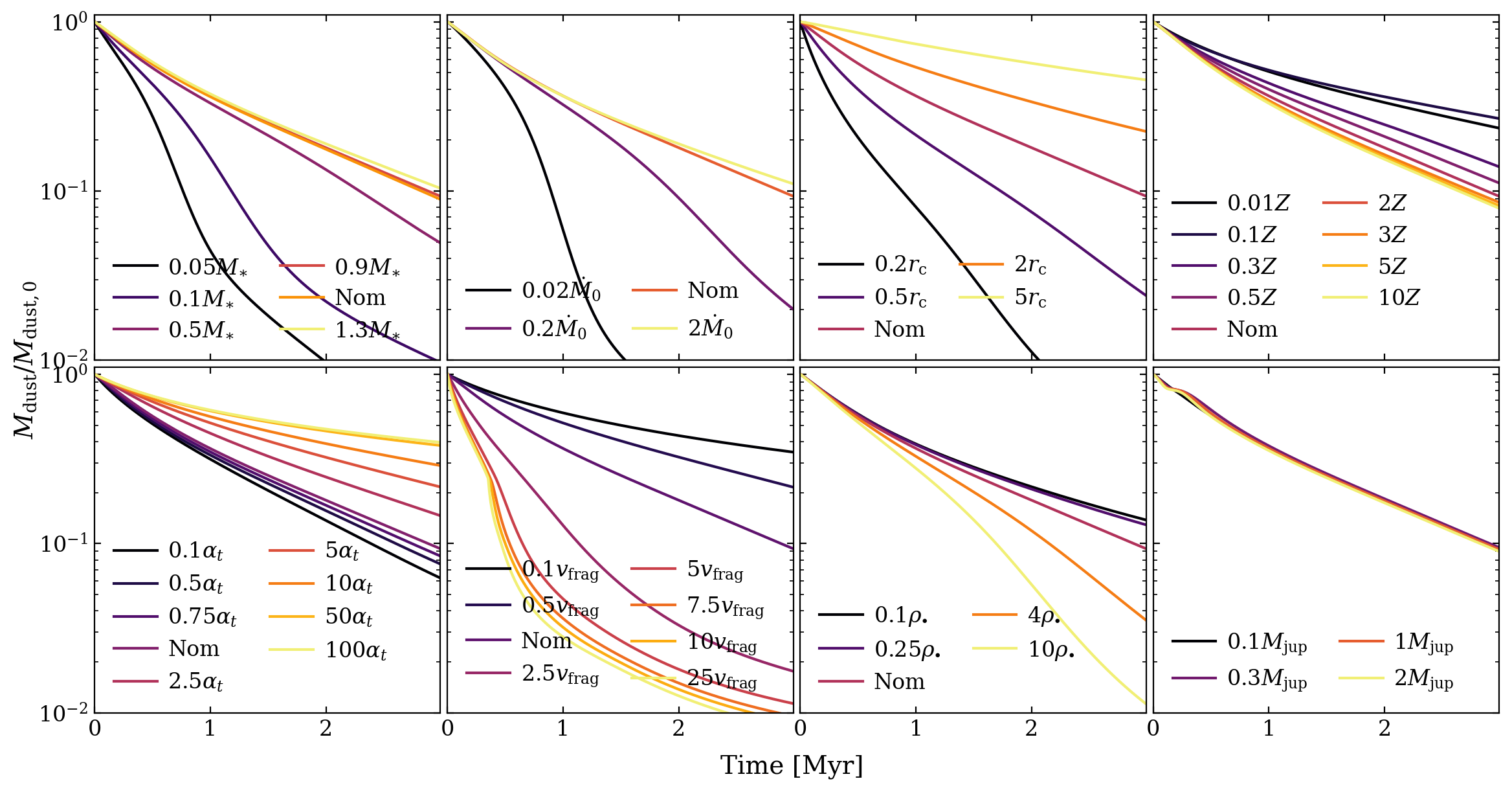}
    \caption{Same as Figure \ref{fig: Mdust vs time}, but normalized to the total dust mass at $t=0$.}
    \label{fig: normalized total dust mass}
\end{figure*}

\begin{figure}
    \centering
    \includegraphics[width=\linewidth]{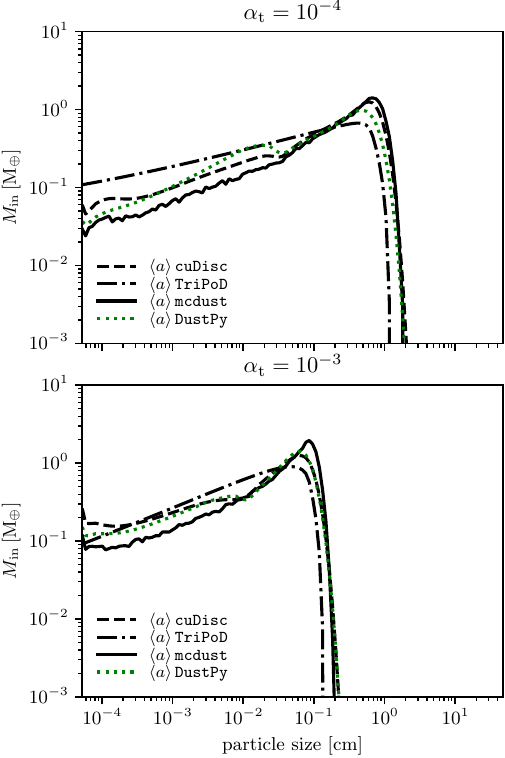}
    \caption{Dust mass distribution in Earth masses in the inner disc ($r<15\, \mathrm{au}$), for our two-dimensional simulations with weak turbulence (top) and strong turbulence (bottom).}
    \label{fig:InnerDiskMassDistr2D.pdf}
\end{figure}

\autoref{fig: st dist param} shows the Stokes number distribution for the same simulations that are shown in \autoref{fig: size dist param}. \autoref{fig: normalized total dust mass} displays the same data as \autoref{fig: Mdust vs time}, but the dust mass has been normalized to the initial dust mass. 
We show the total dust mass distribution within $15\, \mathrm{au}$ in \autoref{fig:InnerDiskMassDistr2D.pdf} for the low and high turbulence cases.

\section{Numerical and physical limits of the dust evolution codes}
During our study, we identified several cases where one or more of the 1D radial simulations yielded nonphysical results. We summarize these cases below:

\begin{itemize}
    \item \textbf{High $v_{\rm frag}$:} When $v_{\rm frag} \gtrsim 5\, \textrm{m\,s}^{-1}$, an inward-moving wave is generated at the outer boundary due to low gas densities and outward-diffusing particles. Using \texttt{adp=False} and a large semimajor axis grid makes this effect less pronounced but does not eliminate it. In the case of \texttt{tripod}, a second inward-moving wave is produced that is not seen in \texttt{DustPy} or \texttt{two-pop-py}. In \texttt{two-pop-py}, the solution also becomes unstable in the Stokes-I regime, corresponding to the innermost regions of the disc. If fragmentation velocities as high as $50\, \textrm{m\,s}^{-1}$ are used, the results can no longer be considered reliable, as the underlying physics was not derived for such large particles. In this case, an unphysical transition is also observed in the \texttt{tripod} simulations. When a Jupiter-mass planet is included in the simulation, nonphysical results are observed already for $v_{\rm frag}>5\, \textrm{m\,s}^{-1}$.
    
    \item \textbf{Deep planetary gaps:} When considering a planetary mass of three Jupiter masses, the density gradient at the gap edge became very steep, reducing the simulation timestep significantly. As a result, the simulation could not be evolved over a full disc lifetime. 

    \item \textbf{Small minimum dust size:} If the minimum size on the dust grid in \texttt{DustPy} is very small, the simulation runtime sometimes becomes extremely long. We encountered this when using $a_0 = 10^{-5}\, \textrm{cm}$ and a minimum dust size of $5\times10^{-6}\, \textrm{cm}$, in combination with $v_{\rm frag} = 0.1\, \textrm{m\,s}^{-1}$ or $\rho_{\bullet} = 0.1\, \textrm{g\,cm}^{-3}$. 

    \item \textbf{Large minimum dust size:} If the minimum size on the dust grid is too large, the size distribution sometimes shows an unphysical increase toward the smallest sizes.
\end{itemize}


\bsp	
\label{lastpage}
\end{document}